\documentclass[11pt]{article}

  \usepackage{amsthm}
  \usepackage{amssymb}
  \usepackage{amsmath}

\newcommand{\K}{\mathcal{K}} 
\newcommand{\Mk}{\mathbb{M}^4} 

\newcommand{\sst}[1]{\scriptscriptstyle{#1}}

\newcommand{\A}{\mathfrak{A}}
\newcommand{\Ob}{\mathfrak{R}}

\newcommand{\si}{\sigma}

\newcommand{\ga}{\gamma}
\newcommand{\be}{\beta}
\newcommand{\te}{\theta}
\newcommand{\eps}{\varepsilon}
\newcommand{\io}{\iota}
\newcommand{\al}{\alpha}
\newcommand{\la}{\lambda}
\newcommand{\Dt}{\Delta_{\mathrm{t}}}
\newcommand{\rhoB}{ \hat{\rho} }
\newcommand{\siB}{ \hat{\si} }

\newcommand{\gaB}{ \hat{\ga} }

\newcommand{\varphiB}{ \hat{\varphi} }
\newcommand{\Bim}{\mathcal{B}(\boldsymbol{\A})}
\newcommand{\Pres}{\mathcal{B}(\boldsymbol{\A^{\sst{\perp}}})}

\newcommand {\defi}{\equiv}
\newcommand {\norm}[1]{\Vert{#1}\Vert}

\author{Giuseppe Ruzzi \\
 \small{Dipartimento di Matematica, Universit\`a di Roma ``Tor Vergata'' }\\
    \small{Via della Ricerca Scientifica I-00133, Roma,  Italy}  \\
          \small{\texttt{ruzzi@mat.uniroma2.it}}}

\title{Essential Properties of the Vacuum Sector for a
                    Theory of Superselection Sectors}

\begin{document}
  \maketitle

\begin{abstract}
As a generalization of DHR analysis, the superselection
sectors are studied in the absence of the spectrum
condition for the reference representation.
Considering a net of local observables in
4-dimensional Minkowski spacetime,
we associate to a set of
representations, that are local excitations of a reference representation
fulfilling Haag duality, a symmetric tensor $\mathrm{C}^*$-category
$\Bim$ of bimodules of the net, with subobjects and direct sums.
The existence of conjugates
is studied  introducing an equivalent formulation of the theory
in terms of  the presheaf associated with the observable net.
This allows us to find, under the assumption
that the local algebras in the reference representation
are properly infinite,
necessary and sufficient conditions for the existence
of conjugates. Moreover, we present several results that suggest
how the mentioned assumption on the reference representation
can be considered essential also in the case of theories in curved spacetimes.
\end{abstract}

\newtheoremstyle{break}
  {}
  {}
  {\em}
  {}
  {\bfseries}
  {.}
  {0.3em}
  {}

\newtheoremstyle{ciccio}
  {}
  {9pt}
  {\rm}
  {}
  {\bfseries}
  {.}
  {0.3em}
  {}

   \theoremstyle{break}
   \newtheorem{teo}{Theorem}[section]
   \newtheorem{prop}[teo]{Proposition}
   \newtheorem{cor}[teo]{Corollary}
   \newtheorem{lemma}[teo]{Lemma}

  \theoremstyle{ciccio}
   \newtheorem{es}[teo]{Example}
   \newtheorem*{oss}{Remark}
   \newtheorem{df}[teo]{Definition}


\tableofcontents
\markboth{Contents}{Contents}

\section{Introduction}
        \label{int}

In a series  of papers \cite{DHR3,DHR4,DR2}, the first two of which
are known as DHR analysis,
Doplicher, Haag and Roberts have shown that the properties
of charges associated with a global gauge group, like
the Bose-Fermi alternative and the charge conjugation symmetry,
find a natural description in the superselection sectors of a net of
local observables. The theory was based on
one important result
obtained in a previous  investigation \cite{DHR1}:
the representations  of the net local observables,
corresponding to such kind of charges, fulfill the
following property: they are 
\textit{local excitations} of the \textit{vacuum representation}.
This property was used in \cite{DHR3, DHR4} as the criterion
for selecting a set of representations of a net of local observables.
The authors associate to this set  a
$\mathrm{C}^*$-category, in which the charge structure arises
from the existence of a tensor product, a symmetry and
a conjugation. Finally it has been shown by
Doplicher and Roberts \cite{DR2} that the unobservable
quantities underlying the theory, namely  the fields and the
global gauge group,  can be reconstructed from the observables. \\
\indent At the present time it is not possible to apply this program
in a curved spacetime without a global symmetry.
In this case, in fact, a notion
corresponding to the spectrum condition by which one could define
a vacuum representation of a net of local observables does not exist
yet\footnote{The superselection sectors
of a net of local observables on an arbitrary globally
hyperbolic spacetime have been studied in \cite{GLRV}.
Except when geometrical obstructions
are present, the results of the DHR analysis are reproduced. However,
the reference representation used in this analysis
is not characterized by physical conditions,
as the vacuum in the case of Minkowski space,
but only by  mathematical ones, suggested by a study on
the representations,  induced by quasi-free Hadamard states,
of the local algebras of a free Bose field \cite{Ve}.
In this connection see also \cite{Ve1}.}. The DHR analysis is,
however, well suited for  treating this situation,
because no explicit use of Poincar\'e covariance is made.
Moreover, the spectrum condition is not fully used  in the theory: 
only the Borchers property, a consequence of the
spectrum condition \cite{Bo}, has a real role.
In this paper we further generalize the theory. We will consider the set
of representations  that are  local excitations of a reference
representation, which is not required to satisfy 
the Borchers property.
Also in this case, a tensor $\mathrm{C}^*$-category
having a symmetry is associated with this set of representations.
Then, the subject of this paper will be the search for
a  criterion for selecting
the \textit{relevant}  subcategory of the theory:
namely, the maximal full subcategory which is closed
under tensor products, direct sums and subobjects,
and whose objects have conjugates.\\[5pt]
\indent In the usual setting of Algebraic Quantum Field Theory
 (see \cite{Ha} and references therein),
we consider a local net of von Neumann algebras
$\boldsymbol{\Ob}$ over $\Mk$, namely a correspondence
$\boldsymbol{\Ob}:\K\ni a \longrightarrow \Ob(a)$
associating to an open  double cone $a$ of $\Mk$ a von Neumann
algebra $\Ob(a)$ on a fixed Hilbert space $\mathcal{H}$,
subject to the conditions:
\[
\begin{array}{lr}
 \qquad a_1\subseteq a_2 \ \ \Rightarrow \ \ \Ob(a_1)\subseteq\Ob(a_2)
 &  \qquad isotony \\
 \qquad a_1\perp a_2 \ \ \Rightarrow \ \ \Ob(a_1)\subseteq\Ob(a_2)'
 & \qquad locality
\end{array}
\]
where the symbol $\perp$ stands for
 spacelike separation   and the prime
for the commutant. The algebra  $\Ob(a)$ is 
generated by all the observables measurable within $a$.
For an unbounded region $S\subseteq\Mk$ there is an associated
$\mathrm{C}^*$-algebra $\Ob(S)$ generated by all the
algebras $\Ob(a)$ such that $a\in\K, \ a\subset S$. We denote by
$\Ob$ the algebra associated with $\Mk$.
As \textit{reference  representation} of $\Ob$ we consider
a locally normal, faithful, irreducible representation $\pi_o$,
on an infinite dimensional separable Hilbert space $\mathcal{H}_o$, 
such that the net of local
observables in the reference representation
\[
\boldsymbol{\A}:\K\ni a\longrightarrow \A(a) \subset \A
\]
where  $\A(a)\defi\pi_o(\Ob(a))$ and  $\A\defi\pi_o(\Ob)$,
satisfies \textit{Haag duality}, namely
\[
   \qquad \A(a^{\sst{\perp}})' = \A(a) \qquad   \qquad \forall a\in\K
\]
where $a^{\sst{\perp}}$ denotes the spacelike complement  of $a$.
Now, in the present investigation we are interested
in a set of representations of $\Ob$ which is closed under
direct sums and subrepresentations, and whose elements are
local excitations of $\pi_o$.
Without the Borchers property,
such a set of representations  can  be selected by a suitable generalization
of the DHR criterion.
Precisely,
we consider the representations $\pi$ of $\Ob$
satisfying the following relation:
for each $a\in\K$  there exists $n_a\in\mathbb{N}$ and an isometry
$V_a :\mathcal{H}_{\pi} \longrightarrow
     \mathcal{H}_o\otimes\mathbb{C}^{n_a}$
such that
\begin{equation}
\label{int:2}
\qquad V_a \cdot \pi(A) \ = \ \pi_o(A)\otimes 1_{n_a} \cdot V_a
                 \ \qquad A\in\Ob(a^{\sst{\perp}})
\end{equation}
We denote by $\mathrm{SC}$ the set of the representations verifying
this selection criterion.\\[5pt]
\indent We will associate to $\mathrm{SC}$ the
tensor $\mathrm{C}^*$-category  $\Bim$ of the
localized transportable bimodules
of the net. This category is closed under direct sums and subobjects.
Moreover, we will show the existence of a symmetry $\eps$, thus,
a notion of statistics of sectors can be introduced.
However, since there might exist objects without left inverses, not
all the sectors of $\Bim$ fall into
the DHR classes of sectors with finite/infinite statistics.
The study of the properties of objects having conjugates
will provide that, apart from the finiteness of the statistics,
an additional condition, called \textit{homogeneity}, is necessary
for the existence of conjugates. Under the
assumption that the local algebras are properly infinite,
we will prove that the homogeneous sectors with finite statistics
have conjugates.\\[5pt]
\indent The key  result that will allow us to formulate the property
of a homogeneous object  is that  the superselection
sectors theory of the net $\boldsymbol{\A}$
is equivalent to the one of the presheaf
$\boldsymbol{\A^{\sst{\perp}}}:\K\ni a\longrightarrow \A(a)'$.
Namely, we will introduce the category  $\Pres$ of the localized
transportable bimodules of the presheaf $\boldsymbol{\A^{\sst{\perp}}}$:
a bimodule $\hat{\rho}$ of $\boldsymbol{\A^{\sst{\perp}}}$
 is a collection of morphisms
\[
_a\rho:\A(a)'\longrightarrow\mathfrak{B}
(\mathcal{H}_o)\otimes \mathbb{M}_{n_\rho} \qquad  \forall a\in\K
\]
compatible  with the presheaf structure.
We will show that this category is isomorphic to $\Bim$;
in particular, any object
$\rho$ of $\Bim$ admits a canonical
extension to a localized transportable
bimodule $\hat{\rho}$ of the presheaf
(Theorem \ref{Ab:3}).
Using this isomorphism, we introduce the notion of
\textit{presheaf-left inverse} of  $\rho$ which
generalizes the concept of left inverse for unital endomorphisms of a
$\mathrm{C}^*$-algebra to its extension $\hat{\rho}$ to the presheaf
(Definition \ref{Cb:2}). However,
the property of admitting  presheaf-left inverses
is not stable under equivalence and depends on the double cone
where the object is localized. Hence, we will say
that $\rho$ is homogeneous whenever \textit{all} the elements
of its equivalence class $[\rho]$ admit presheaf-left inverses
(Definition \ref{Cb:2a}).\\[5pt]
\indent The existence of a maximal full
subcategory $\Bim_\mathrm{fh}$ of
$\Bim$ with homogeneous objects,
closed under direct sums, tensor products and subobjects,
and having finite statistics, will be proved in Proposition \ref{Eb:2}.
Any object of $\Bim_\mathrm{fh}$ is a finite direct sum of irreducible objects
$\rho$ fulfilling the following conditions:
there exists an integer $d$, an object $\ga$
and an isometry $V\in(\ga,\rho^{\sst{d}})$ such that
\begin{enumerate}
\item $\eps(\ga,\ga) = \pm 1_{\ga}$;
\item $VV^+$ equals either $\mathrm{A}_{d}$ or $\mathrm{S}_d$
       the totally (anti)symmetric projector
in $(\rho^{\sst{d}},\rho^{\sst{d}})$;
\item the extension $\hat{\ga}$ of $\ga$
       is a faithful morphism of the presheaf, that is
        $\phantom{k}_a\ga:\A(a)'\longrightarrow
    \mathfrak{B}(\mathcal{H}_o)\otimes\mathbb{M}_{n_\rho}$ is a
       faithful morphism for each $a\in\K$.
\end{enumerate}
$\Bim_\mathrm{fh}$  is the  \textit{relevant} subcategory
of the theory. Indeed, we will prove that
on the one hand  each object with conjugates belongs to this category
(Theorem \ref{F:1}), and, on the other hand, if the local algebras
are properly infinite any object of $\Bim_\mathrm{fh}$ has conjugates
(Theorem \ref{F:2}). This last result
suggests that \textit{it is reasonable to
include  proper  infiniteness  of the local algebras $\A(a)$ as an axiom of
the theory}. This proposal is also supported
by the following facts: first, this property can be derived,
in a particular case, from the existence
of conjugates (Theorem \ref{F:3});  secondly, in a globally hyperbolic
spacetime
the algebras of local observables of a multiplet of $n$ Klein-Gordon
fields in any Fock representation, acted on by $U(n)$ as a global gauge
group, fulfill this property (this result is proved in \cite{Ru} and
it will be described in a forthcoming article).\\[5pt]
\indent The paper is organized as follows: in Section \ref{A} we
introduce the categories $\Bim$ and $\Pres$ and show that they are
isomorphic; Section \ref{AB} is entirely devoted to the construction
of the category  $\Bim_\mathrm{fh}$; 
in Section \ref{F} we study the conjugation
and derive the above stated solutions;
finally, Section \ref{G} concludes the work. In Appendix \ref{X}
some definitions and results on tensor $\mathrm{C}^*$-categories 
are presented.
\section{The net and the presheaf approach to the theory,
and their equivalence}
\label{A}
In this section we introduce the categories
$\Bim$  and $\Pres$ which are respectively  the categories
of  localized transportable bimodules of the net and
of the presheaf. We show how these categories are related to
$\mathrm{SC}$ and that they are isomorphic\footnote{The
relevance of  sheaves of von Neumann algebras
in the theory of superselection sectors was pointed out
for the first time by J.E. Roberts \cite{Ro3} who showed
a correspondence between sectors and some  Hermitian bimodules
over a sheaf of von Neumann algebras on Minkowski space.}.
We conclude  by introducing  the notions of faithfulness and
of double faithfulness for the objects of $\Bim$.
\subsection{The category $\Bim$}
\label{Ao}
We show that there is, up to unitary equivalence,
a bijective correspondence  between $\mathrm{SC}$ and the set $\Dt$
of the localized transportable morphisms of the net.
After the observation that the elements of $\Dt$ are bimodules of
$\A$, the category $\Bim$ is defined as the category whose set of objects
is $\Dt$ and whose arrows are the intertwiners of the elements of $\Dt$.\\[5pt]
\indent A morphism $\rho:\A\longrightarrow
\mathfrak{B}(\mathcal{H}_o)\otimes\mathbb{M}_{n_\rho}$
is said to have \textit{multiplicity $n_\rho$} and to be
\textit{localized in}  $o$ if for any $a\in\K, \ a\perp o$ then
\[
\rho(A) = A\otimes 1_{n_\rho} \cdot \rho(1) \qquad \forall A\in \A(a).
\]
We denote by $\Delta$  the set of localized morphisms
and by $\Delta(o)$ the subset of those morphisms which are localized
within $o$.
Given $\rho,\si\in\Delta$, the set $(\rho,\si)$  of the
\textit{intertwiners}  between
$\rho$ and $\si$ is the set of the operators
$T\in\mathfrak{B}(\mathcal{H}_o\otimes \mathbb{C}^{n_\rho},
    \mathcal{H}_o\otimes \mathbb{C}^{n_\si})$ such that

\[
 T\rho(1) = T = \si(1)T  \qquad \mbox{and}
\qquad T\rho(A)=\si(A)T \ \ \   \forall A\in\A
\]
A localized morphism $\rho$ is said to be \textit{transportable} if for each
$o$ there exists $\tau\in\Delta(o)$ and a unitary
$U\in(\rho,\tau)$. We denote by  $\Dt$ the set of
localized transportable morphisms and by $\Dt(o)$ the subset
of those morphisms  which are localized in $o$. \\
\indent The following lemma is an
easy consequence of Haag duality and of the localization property
of the elements of $\Dt$.
\begin{lemma}
\label{Ao:2}
Let $\rho\in\Dt(o)$. Then the following assertions hold: \\
a) for each  $a\in\K, \ o\subseteq a $ we have
$\rho(\A(a))\subseteq \A(a)\otimes\mathbb{M}_{n_\rho}$;  \\
b) if $\si\in\Dt(o_1)$ and $T\in(\rho,\si)$, then $T$ has values
$T_{i,j}$ in $\A(a)$
$\forall \ a\in\K, \ o\cup o_1 \subseteq a$.
\end{lemma}
Studying $\mathrm{SC}$ is equivalent to studying  $\Dt$ because
there exists, up to unitary equivalence,  a bijective correspondence
between the representations satisfying $\mathrm{SC}$ and the
morphisms in $\Dt$. In fact,
for  each $\pi\in\mathrm{SC}$ there is a  corresponding
set of localized transportable morphisms defined
as follows
\[
\rho_a(A)\defi V^+_a\pi_o(A)V_a \qquad A\in\A,  \ a\in\K
\]
where $\{V_a\}_{a\in\K}$ is  a set of isometries associated with
$\pi$ by (\ref{int:2}). Conversely, given $\rho\in\Dt$
then
\[
\qquad \pi(A) \defi \rho(\pi^{\sst{-1}}_o(A))\qquad A\in\Ob
\]
is a representation belonging to $\mathrm{SC}$. \\[5pt]
\indent In order to associate a
category with $\Dt$, and hence with $\mathrm{SC}$, we need to introduce
the tensor $\mathrm{C}^*$-category $\mathcal{B}$
\textit{of bimodules of }the  $\mathrm{C}^*$-algebra $\A$ \cite{DR1}.
The objects are bimodules of $\A$, namely the morphisms
$\rho:\A\longrightarrow \A\otimes\mathbb{M}_{n_\rho}$ with multiplicity
$n_\rho\in\mathbb{N}$. The arrows between
two objects $\rho$, $\tau$ are the intertwiners
$T\in(\rho,\tau)$  with values in $\A$, i.e. $T_{i,j}\in\A$.
The composition law between the arrows is the usual  rows times
columns product, and it is denoted by ``$\cdot$''.
The identity arrow $1_\rho$ of an object $\rho$ is
the projection $\rho(1)$. The adjoint ``$+$'' is defined
as  $\rho^+\defi\rho$ on the objects, and
$T^+_{i,j}\defi T^*_{j,i}$ for each $T\in(\rho,\tau)$,
where $*$ denotes the involution of $\A$.
The  tensor product  ``$\times$'' is defined  by
using the lexicographical ordering. Namely $\times$ is defined on the objects
as
\[
 \rho\si( \ )_{i,j} \defi \rho(\si( \ )_{i_2,j_2})_{i_1,j_1} \qquad \mbox{ where }\qquad
   i = i_1 + n_\rho i_2 \ \ j = j_1 + n_\rho j_2
\]
(observe that $\rho\si$ has multiplicity $n_\rho n_\si$), and
\[
 (T\times S)_{i,j} \defi T_{i_1,k}\rho(S_{i_1,j_2})_{k,j_1}
 \qquad \mbox{ where } \qquad i = i_1 + n_{\rho_2} i_2 \ \ j = j_1 + n_{\rho_1} j_2
\]
for each $T\in (\rho_1,\rho_2)$, $S\in (\si_1,\si_2)$. The identity object $\io$ of
the tensor product is the morphism $\io(A) \defi A$ for each  $A\in\A$.
Since $\A$ is irreducible  $\io$ is irreducible.
Finally, one can easily checks that $\mathcal{B}$ is closed
under direct sums and subobjects. \\[5pt]
\indent Now, returning to the problem of
stating what category is associated
with $\Dt$, we notice that
$\Dt$  is  a  subset of the objects of
$\mathcal{B}$ because of Lemma \ref{Ao:2}.a, and that by Lemma \ref{Ao:2}.b
the set of the intertwiners between $\rho,\si\in\Dt$
is equal to the set of the arrows between $\rho$ and $\si$ as objects
of $\mathcal{B}$. The category $\Bim$ of the
\textit{localized transportable bimodules of} $\boldsymbol{\A}$
is the full subcategory of $\mathcal{B}$ whose objects belong to $\Dt$.
$\Bim$ is closed under tensor products,
direct sums and  subobjects, and the identity object $\io$ is
irreducible. In conclusion,  $\Bim$ is the category  associated with  $\Dt$
that we were  looking for.  The
\textit{superselection sectors of the theory} are the unitary
equivalence classes of the irreducible objects  of $\Bim$.
\subsection{The category $\Pres$}
\label{Aa}
The presheaf  $\boldsymbol{\A^{\sst{\perp}}}$ associated with the net
$\boldsymbol{\A}$ is defined as the correspondence
\[
\boldsymbol{\A^{\sst{\perp}}}:\K\ni a\longrightarrow \A(a)'
\]
where for $a\subseteq b$  the restriction
$\A(b)'\longrightarrow\A(a)'$   is given by the
inclusion $\A(b)'\subseteq \A(a)'$.
A \textit{morphism} $\rhoB$ of $\boldsymbol{\A^{\sst{\perp}}}$ is a collection
$ _a\rho:\A(a)'\longrightarrow
\mathfrak{B}(\mathcal{H}_o)\otimes \mathbb{M}_{n_\rho}$, $a\in\K$,
of morphisms with a fixed multiplicity $n_\rho$, fulfilling the relations:
\[
\begin{array}{llr}
 a) & _a\rho(1) \ = \ \rho(1) \ \ \ \ \ \forall a\in\K \\
 b) & \mbox{if } \ \ a \subseteq b \ \mbox{ then } \
      _a\rho \upharpoonright \A(b)' \ = \ _b\rho & \qquad (compatibility)
\end{array}
\]
In a way similar as has been done for the net, the notion
of localized transportable morphism of the presheaf can be
introduced.
A morphism $\rhoB$ of  $\boldsymbol{\A^{\sst{\perp}}}$ is said to be
\textit{localized} in $o$ if
\[
_o\rho(A) \ = \ A\otimes 1_{n_\rho}\cdot \rho(1) \qquad\forall A\in\A(o)'
\]
We denote by $\Delta^{\sst{\perp}}$  the set of  localized morphisms
and by $\Delta^{\sst{\perp}}(o)$ the subset of those morphisms localized
within $o$.
Given $\rhoB,\siB\in\Delta^{\sst{\perp}}$, the set $(\rhoB,\siB)$ of the
\textit{intertwiners} between $\rhoB$ and $\siB$ is
the set of the operators
$T\in\mathfrak{B}(\mathcal{H}_o\otimes \mathbb{C}^{n_\rho},
\mathcal{H}_o\otimes \mathbb{C}^{n_\si})$ such that
\[
  T\rho(1)  =  T =   \si(1)T  \qquad  \mbox{and} \qquad
    T  {_a\rho}(A)  =   {_a\si}(A)T  \ \ \ \
       \forall A\in\A(a)', \ \forall a\in\K
\]
A localized morphism $\rhoB$  is said to be
\textit{transportable} if for each $a\in\K$
there exists $\siB\in\Delta^{\sst{\perp}}(a)$ and a unitary
$U\in(\rhoB,\siB)$. By $\Dt^{\sst{\perp}}$ we denote the set of
localized transportable morphisms  and by $\Dt^{\sst{\perp}}(o)$
the subset of transportable morphisms which  are localized within $o$.
Finally, we call the
\textit{category of localized transportable bimodules
of $\boldsymbol{\A^{\sst{\perp}}}$}, and denote it by $\Pres$,
the category whose set of objects  is
$\Dt^{\sst{\perp}}$, and whose  set of arrows between $\rhoB,\siB\in\Dt^{\sst{\perp}}$
is $(\rhoB,\siB)$. Clearly, $\Pres$
is a $\mathrm{C}^*$-category closed under direct sums and subobjects.
\begin{prop}
\label{Aa:2}
Let $\rhoB\in\Dt^{\sst{\perp}}$ be localized in $o$. \\
a) For each $a,b,c\in\K, \ c\perp a, b$ we have \
$ \  _a\rho\upharpoonright  \A(c) \ =
       \  _b\rho  \upharpoonright   \A(c)$ \\
b) If $c\in\K, \  c\perp o$, then \
$_a\rho(A)  \ = A\otimes 1_{n_\rho}\cdot \ \rho(1)$ \ \
          $\forall A\in\A(c)$ and $\forall a\in\K, \ a\perp c$. \\
c) $_a\rho(\A(a)')\subseteq \A(a)'\otimes\mathbb{M}_{n_\rho}$
   \  for each  $a\in\K, \ o\perp a$. \\
d) Given  $\siB\in\Dt^{\sst{\perp}}(b)$  and  $T\in(\rho,\si)$
  then $T_{i,j}\in \A(a)$ for each $a\in\K, \ o\cup b\subseteq a$.
\end{prop}
\begin{proof}
a) Since the  spacelike complement  of a double cone
is pathwise connected
in  $\Mk$,
there is a path $p$, contained in $c^{\sst{\perp}}$, joining $a$ to $b$.
As $\A(c)$ is contained in the commutant of the algebras associated
with each double cone of the path, the proof follows from the
compatibility of the morphisms.
b) Notice that  $c\perp  o, a$. Then,  by a) we have
  $_a\rho(A)= {_o\rho}(A)$ for each $A\in\A(c)$.
Since $\rhoB$ is localized in $o$,  the proof follows
 from the fact that  $\A(c)\subset\A(o)'$. c) is postponed to
the next section.
d) follows from b).
\end{proof}
Some comments are in order. First,
the proposition does not hold in a 2-dimensional
Minkowski spacetime because the spacelike complement
of a double cone
is not pathwise connected.
Secondly, the statement $a)$ does not  depend on  the double cone
where the object is localized. Thirdly, notice that, once
$o\in\K$ is fixed, the correspondence
\begin{equation}
\label{Aa:3}
\{a\in\K \ | \ a\perp o\}
   \ni a \ \longrightarrow  \ \A(a)'
\end{equation}
is a presheaf of von Neumann algebras and,  if $\rhoB\in\Dt^{\sst{\perp}}$
is localized in $o$,  then also the correspondence
\begin{equation}
\label{Aa:4}
\{a\in\K \ | \ a\perp o\}
   \ni
   a \ \longrightarrow  \ (\A(a)'\otimes\mathbb{M}_{n_\rho})_{\rho(1)}
\end{equation}
is a presheaf of von Neumann algebras, because
$\rho(1)\in \A(a)'\otimes\mathbb{M}_{n_\rho}$ for each
$a\in\K$, $a\perp o$; here
$(\A(a)'\otimes\mathbb{M}_{n_\rho})_{\rho(1)}$ denotes the reduced
algebra $ \ \rho(1)(\A(a)'\otimes\mathbb{M}_{n_\rho})\rho(1)$.
Then, as a consequence of Proposition
\ref{Aa:2}.c,
the collection $\{ {_a\rho} \ |  \ a\in\K,  \ a\perp o  \}$
is a presheaf morphism from  (\ref{Aa:3}) to  (\ref{Aa:4}).
In the following we will
refer to the presheaves (\ref{Aa:3}) and  (\ref{Aa:4}) as,
respectively,  the \textit{domain} and the \textit{codomain}  of
$\rhoB$ as  an  element of the set  $\Dt^{\sst{\perp}}(o)$.
\subsection{The isomorphism between $\Pres$ and $\Bim$}
\label{Ab}
The relation between $\Pres$ and $\Bim$ is deeper than the one
suggested from their definition: in fact they are isomorphic.
The key point of the proof consists in proving that each element of $\Dt$
admits an extension to a morphism  of the presheaf. In order to prove
this we need  to introduce the cohomological description
of the theory of superselection sectors
developed by J.E. Roberts \cite{Ro2} (see also \cite{Ro}).
By using the same reasoning of that paper,
it is possible to  introduce the category of 1-cocycles of the net
and show that it is equivalent to $\Bim$. However we limit ourselves
to describing  the way  the set
$Z^1_t(\boldsymbol{\A})$ of 1-cocycles of the net
and $\Dt$ are related.\\[5pt]
\indent Having fixed  a double cone $o$ and $\rho$
localized in $o$, for
each $a\in\K$ let us choose a set of unitary arrows
$V_{ao}\in(\rho,\tau_a)$ where $\tau_a$ is localized in $a$ and
$\rho=\tau_o$. Defining
\begin{equation}
\label{Ab:1}
 z_{ab} \ \defi \ V_{ao}\cdot V^+_{bo} \qquad a,b\in\K .
\end{equation}
and observing that $z_{ab}\in(\tau_b,\tau_a)$, we have the
\textit{1-cocycle identity}
\[
\begin{array}{llll}
&  a) &   z_{ab}\cdot z_{bc} \ = \ z_{ac} \qquad \qquad & a,b,c\in\K\\
\mbox{and that} \ \  &  \\
 & b) &  z_{aa} \ = \tau_a(1) & a\in\K \\
 & c) & z_{ab}  \mbox{ has values in } \A(d) &   a,b,d\in\K,
\ a\cup b\subseteq d
\end{array}
\]
A collection of partial isometries
$\{ \bar{z}_{ab}\}_{a,b\in\K}$ satisfying a)$-$c)
is called a \textit{1-cocycle} of
$\boldsymbol{\A}$. A different choice of the set
$V_{ao} \ a\in\K$ yields a cohomologous  cocycle. Conversely,
we can associate with each
$\bar{z}\in Z^1_t(\boldsymbol{\A})$ an element of $\Dt$.
Note that for each $a,b\in\K$, $a\perp b$ we have
\begin{equation}
\label{Ab:2}
\rho(A)= z_{ob}\cdot \tau_b(A)\cdot z_{bo}= z_{ob} \cdot A \otimes 1_{n_b}
\cdot z_{bo} \qquad A\in\A(a).
\end{equation}
Then by replacing $z$ with $\bar{z}$ in the r.h.s.
of (\ref{Ab:2}) one gets a transportable morphism localized in $o$.
\begin{teo}
\label{Ab:3}
The categories $\Pres$ and $\Bim$ are isomorphic.
\end{teo}
\begin{proof}
First we define  the \textit{extension} functor
$\mathrm{E}:\Bim\longrightarrow\Pres$.
Let $\rho\in\Dt$ be localized in $o$, and let $z$ be the 1-cocycle
defined by (\ref{Ab:1}). We set
\[
\begin{array}{rcll}
_a\mathrm{E}(\rho)(A)  & \defi & z_{oa}\cdot A\otimes 1_{n_a}\cdot z_{ao} &
        \ \ \ A\in\A(a)' \\
     \mathrm{E}(T) & \defi &\ T & \ \ \ T\in(\rho,\si)
\end{array}
\]
For each $a\in\K$, $_a\mathrm{E}(\rho)$ is a normal morphism of
$\A(a)'$  and   $_a\mathrm{E}(\rho)(1) = \rho(1)$.
If $a\subseteq b$ then for each $A\in\A(b)'$ we have:
\[
_a\mathrm{E}(\rho)(A) \  =  \  z_{oa} \ A\otimes 1_{n_a} \ z_{ao}
         \ =  \ z_{ob}z_{ba}\cdot
        \ A\otimes 1_{n_a} \cdot
        z_{ba}z_{bo}
         \  = \  _b\mathrm{E}(\rho)(A)
\]
because the coefficients of $z_{ba}$ belong to $\A(b)$. Moreover,
$_o\mathrm{E}(\rho)(B)  =  B\otimes 1_{n_o}\cdot \rho(1)$ for each
$B\in\A(o)'$  because $z_{oo}=\rho(1)$.
Hence $\mathrm{E}(\rho)$ is a morphism of the presheaf and it is
localized in $o$. It is worth  observing that, by  (\ref{Ab:2}), we have
\[
_a\mathrm{E}(\rho)(A) =
z_{oa} \cdot A\otimes 1_{n_a} \cdot z_{ao}  = \rho(A)\qquad A\in\A(a^{\sst{\perp}}).
\]
Thus, $_a\mathrm{E}(\rho)$ is a normal extension of $\rho$ to
$\A(a)'$ and  it is unique because, by Haag duality, $\A(a^{\sst{\perp}})$
is weakly dense in $\A(a)'$. After this observation it is easy to see
that $T\in(\mathrm{E}(\rho),\mathrm{E}(\si))$ and consequently
that  $\mathrm{E}(\rho)$ is transportable. \\
\indent We now pass to define the \textit{restriction}
functor  $\mathrm{R}: \Pres \longrightarrow \Bim $.
Let $\rhoB\in\Dt^{\sst{\perp}}$.  Given $a\in\K$, we take
$b\in\K , \ a\subset b^{\sst{\perp}}$
and define
\[
\begin{array}{rcll}
 \mathrm{R}(\rhoB)(A) &  =  &  _b\rho(A) \qquad &  A\in\A(a)\\
 \mathrm{R}(S)  &  \defi &  S & S\in(\rhoB,\siB)
\end{array}
\]
$\mathrm{R}(\rhoB)$ is a morphism
of $\A(a)$ and $\mathrm{R}(\rhoB)(1) = \rho(1)$.
By the Proposition \ref{Aa:2}.a
$\mathrm{R}(\rhoB)$ does not depends on the choice
of $b \perp a$ and, for this reason, it is compatible
with the net $\boldsymbol{\A}$. Hence it is extendible by continuity
to a morphism of $\A$.
If $\rhoB$ is localized in $o$ then, by
Proposition \ref{Aa:2}.b, $\mathrm{E}(\rhoB)$ is also localized in
$o$. The proofs both that $\mathrm{R}(\rhoB)$ is transportable and
that $S$ belongs to $(\mathrm{R}(\rhoB),\mathrm{R}(\siB))$ are
straightforward and,
therefore, we omit them. Finally,
observing that $\mathrm{R}(\rhoB)$
is the restriction of the components of $\rhoB$ to the algebras
of double cones, it easily follows that
$\mathrm{R}\circ\mathrm{E} = id_{\Bim}$ and
$\mathrm{E}\circ\mathrm{R} = id_{\Pres}$.
\end{proof}
Concerning the functors $\mathrm{E}$ and $\mathrm{R}$
introduced in the previous theorem,
from now on we will use the following notation:
we will denote by $\rhoB$
the extension $\mathrm{E}(\rho)$ of $\rho\in\Dt$;
conversely, we will denote by $\si$ the restriction $\mathrm{R}(\siB)$
of $\siB\in\Dt^{\sst{\perp}}$.\\[5pt]
\indent As a first consequence of Theorem \ref{Ab:3}, we prove
Proposition \ref{Aa:2}.c. Given $\rhoB\in\Dt^{\sst{\perp}}(o)$,
let $z$ be the 1-cocycle, defined by (\ref{Ab:1}), associated with $\rho$.
Let $a\in\K, \ a\perp o$. Then for each
$A\in\A(a)$ and $B\in\A(a)'$ we have $_a\rho(B)\cdot A\otimes 1_{n_o}$
$=z_{oa}\cdot B\otimes 1_{n_a} \cdot z_{ao}\cdot A\otimes 1_{n_o}$
$ = z_{oa}\cdot B\otimes 1_{n_a}\cdot \tau_a(A) \cdot z_{ao}$
$=z_{oa}\cdot \tau_a(A) \cdot B\otimes 1_{n_a} \cdot z_{ao}$
$=A\cdot {_a\rho}(B)$, where the inclusion
$\tau_a(\A(a))\subseteq\A(a)\otimes\mathbb{M}_{n_a}$ has been used. This
completes the proof.\\[5pt]
\indent Secondly, a tensor product can be easily introduced on $\Pres$:
\[
\begin{array}{ccl}
 \Dt^{\sst{\perp}} \ni\rhoB,\siB\longrightarrow
   \rhoB \boxtimes  \siB\in\Dt^{\sst{\perp}}  \ \ \ \ &  \mbox{ where } &
   \rhoB \boxtimes \siB \defi \  \widehat{\rho\si} \\
 (\rhoB_1,\rhoB_2) , (\siB_1,\siB_2)   \ni   T , S
   \longrightarrow   T  \boxtimes  S
\in(\rhoB_1  \boxtimes  \siB_1, \rhoB_2  \boxtimes  \siB_2) & 
\mbox{ where } &
 T  \boxtimes  S \defi \ T\times S
\end{array}
\]
A useful property of $\boxtimes$ is shown in the following
\begin{prop}
\label{Ab:5}
Let $\rho\in\Dt(a)$, $\si\in\Dt(b)$. If
$c\in\K, \ a\cup b  \perp c$ then
\[
_c(\rho  \boxtimes  \si)(A)  \ = \ _c\rho({_c\si}(A))\qquad
 A\in\A(c)'
\]
\end{prop}
\begin{proof}
Without loss of generality we prove the statement  only
in the case of objects with multiplicity equal to one
(namely endomorphisms, in general not unital, of the algebras $\A(a)'$).
Let $z,\bar{z}$ be two 1-cocycles
associated with $\rho$ and $\si$ respectively. First we observe that
$_c(\rho  \boxtimes  \si)(A) \ = \
  {_c(\rho\si)}(A) \ = \ z_{oc}\times \bar{z}_{bc} \cdot  A \cdot
z_{co}\times \bar{z}_{cb}$. Taking $d,e,h\in\K$ such that
$d\perp e, \ d\cup e\subset c$, and $b\cup e \subseteq h, \ h \perp
d$, then for each $A\in\A(c)'$  we have
\begin{multline*}
_c(\rho  \boxtimes \si)(A)
  = z_{od}z_{dc}\times  \bar{z}_{be}\bar{z}_{ec}  \cdot
  A\cdot  z_{cd}z_{do} \times \bar{z}_{ce}\bar{z}_{eb} \\
  = z_{od}\times  \bar{z}_{be}\cdot z_{dc}\times  \bar{z}_{ec}  \cdot
  A\cdot  z_{cd}\times \bar{z}_{ce}\cdot z_{do} \times \bar{z}_{eb}
    = z_{od}\times \bar{z}_{be}  \cdot A\cdot  z_{do} \times \bar{z}_{eb}
\end{multline*}
where the last equality holds because $z_{dc}\times \bar{z}_{ec}\in\A(c)'$.
Observing that
$ z_{od}\times \bar{z}_{be} = z_{od}\cdot \tau_d(\bar{z}_{be}) =  z_{od}\cdot
\bar{z}_{be}$,  because $\bar{z}_{be}\in\A(h)$, and that,
by Proposition \ref{Aa:2}.c,   ${_c\si}(\A(c)')\subseteq \A(c)'$,
then it follows that
$z_{od}\times \bar{z}_{be}  \cdot A\cdot  z_{do} \times \bar{z}_{eb}$
$= z_{od} \ \cdot (\bar{z}_{be}
    \cdot A\cdot   \bar{z}_{eb}) \cdot z_{do}$
$= {_d\rho}({_e\si}(A)) = {_c\rho}({_c\si}(A)),$
where we used the fact that $\tau_d$ is localized in
$d$. This completes the proof.
\end{proof}
\subsection{Faithfulness and Double Faithfulness}
\label{Ac}
Our aim  is to identify the relevant subcategory of $\Bim$.
A first  step toward the understanding of this problem is made 
in this section.
\begin{df}
\label{Ac:1}
 We say that an object  $\rho$ of $\Bim$ is:
 \begin{itemize}
 \item[i)] \textit{faithful} if it is a faithful morphism of $\A$.
 \item[ii)]\textit{doubly faithful} if its extension $\rhoB$ is a faithful
 morphism of the presheaf, namely, for each $a\in\K$, $_a\rho$ is a
 faithful morphism of the algebra $\A(a)'$.
\end{itemize}
\end{df}
Since  an object $\rho$ of $\Bim$ is the restriction to the local algebras
of its  extension $\rhoB$, double faithfulness
implies faithfulness. The converse is, in general, false
as can be easily seen  by the following proposition.
\begin{prop}
\label{Ac:2}
Let $\rho$ be an object of $\Bim$ and let us denote by $[\rho]$ the
equivalence class of $\rho$. The following assertions hold:\\
a)  $\rho$ is faithful if, and only if,
    for each $o,a\in\K, \ o\perp a$ and for each
    $\tau\in[\rho]$ localized in
    $a$,   the central support of
    $\tau(1)$  in  $\A(o)'\otimes\mathbb{M}_{n_\si}$
    is equal to $1\otimes1_{n_\si}$; \\
b)  $\rho$ is doubly faithful if, and
    only if,  for each $o\in\K$ and   for each $\si\in[\rho]$
    localized in $o$, the
    central support of $\si(1)$ in
    $\A(o)\otimes\mathbb{M}_{n_\si}$ is equal to $1\otimes 1_{n_\si}$.
\end{prop}
\begin{proof}
b) Since the extensions $\rhoB$ and $\siB$ of $\rho$ and $\si$
   are equivalent and $\si$ is localized in $o$,
   for $A\in\A(o)'$ we have $_o\rho(A) =0\iff {_o\si}(A) =
     \si(1)\cdot A\otimes 1_{n_\si}= 0$, and the proof follows
  from the definition of central support. The assertion a)
 follows in a similar way.
\end{proof}
Without any further assumptions on the structure
of local algebras, and in particular on their centers,
we have no way to conclude that these two properties are  fulfilled.
Notice that properties  like  the Schlieder
property\footnote{The net $\boldsymbol{\A}$ has the Schlieder
property  if  given $a,b\in\K, \ a\perp b$  and $A\in\A(a), B\in\A(b)$
then   $A\cdot B=0 \iff  A=0 \mbox{ or } B=0$.} or the simplicity
of $\A$,  which are weaker than
the Borchers property and imply  the faithfulness, cannot
be deduced from the hypotheses we have made on the local
algebras. Thus, we have to accept
the possible existence both of nonfaithful objects
and of not doubly faithful objects.
Since double faithfulness will turn out to be necessary  for the
existence of conjugates, in the following,
not doubly faithful objects shall have to
be excluded from the analysis.\\
\indent Direct sums, tensor products and equivalence
preserve (double) faithfulness:
\begin{prop}
\label{Ac:5}
The following assertions hold: \\
a) if $\rho\in\Dt$ is doubly faithful $\Rightarrow$ each $\si\in[\rho]$
is doubly faithful; \\
b) if $\rho_1,\rho_2\in\Dt$ are doubly faithful $\Rightarrow$
   $\rho_1\oplus\rho_2$ and $\rho_1\rho_2$ are doubly faithful.\\
The same assertions hold for  faithfulness.
\end{prop}
\begin{proof}
$a)$ follows from the fact
that if $\si\in[\rho]$ then $\siB\in [\rhoB]$.
The proof of statement $b)$, concerning  the direct sum,  is obvious.
Given  $a\in\K$, let $b_1,b_2\in\K, \
  b_1,b_2\perp a$  and let $V_i\in(\rho_i,\si_i)$ be unitary
  such that  $\si_i$  is   localized in  $b_i$ for i=1,2.
For $A\in\A(a)'$, by   Proposition \ref{Ab:5},  we have
$  _a(\rho_1 \boxtimes  \rho_2)(A) \ = \
  V^+_1\times V^+_2 \cdot  {_a(\si_1  \boxtimes  \si_2)}(A)
   \cdot V_1\times V_2
    =  V^+_1\times V^+_2 \cdot {_a\si_1} ( {_a\si_2}(A)) \cdot
 V_1\times V_2
$
and the proof is now  completed.
\end{proof}
\section{Statistics and  selection of  the relevant subcategory}
\label{AB}
This section is entirely devoted to showing  how the relevant subcategory
of the theory can be selected. We will find it convenient
to work in the net approach,
that is to say with $\Bim$;
nevertheless, to introduce the notion of a homegenous object
of $\Bim$, we will have to rely on
the presheaf approach. Homogeneity will turn out to be one of the
properties characterizing  the objects of the relevant subcategory.\\
\indent We start by proving the existence
of a symmetry $\eps$. Afterwards, we introduce the notions of net-left
inverse, presheaf-left  inverse and homogeneity.
We prove that each doubly faithful simple object is homogeneous.
After having introduced the category of objects with finite statistics,
we conclude  by showing how the relevant subcategory
can be selected using doubly faithful simple objects.
\subsection{Symmetry}
\label{B}
When a tensor $\mathrm{C}^*$-category  has a symmetry $\eps$
(see definition in appendix) it is possible to introduce,
as in DHR analysis, a notion of statistics of sectors.
Briefly,
one first notes that for each object $\rho$, by means of $\eps$,
there is an  associated  unitary representation  $\eps^n_\rho$ of the
permutation group of $n$-elements $\mathbb{P}(n)$, with values
in $(\rho^{\sst{n}},\rho^{\sst{n}})$.
If $\rho$ is irreducible, \textit{the
statistics of the sector} $[\rho]$ is the collection of the unitary
equivalence classes of the representations
$\eps^n_\rho$ as $n$ varies.
In this section we show that $\Bim$ has a symmetry $\eps$ and,
therefore, an associated  notion of statistics of sectors.\\[5pt]
\indent We start by recalling a result from \cite{DR1}.
Let  $(n,m)$ denote the set of $m\times n$
matrices $A$ with values  $A_{i,j}$ in $\A$. Then, there exists a
map  $\theta:n,m\longrightarrow \theta(n,m)\in(nm,mn)$
with values $\theta(n,m)_{i,j}$ in $\mathbb{C}$, that verifies
\begin{equation}
\label{B:0}
\theta(m,m_1)\cdot A\otimes B \ = \ B\otimes A \cdot \theta(n,n_1)
\end{equation}
for each pair $A\in(n,m)$, $B\in(n_1,m_1)$ with commuting values,
that is to say  $[A_{i,j},B_{l,k}]=0$,
where $(A\otimes B)_{i,j}\defi A_{i_1,j_1}B_{i_2,j_2}$ is the
lexicographical order product.\\[5pt]
\indent The proof of the existence of a symmetry $\eps$ is based on the
following
\begin{lemma}
\label{B:1}
Let $\rho_{i}\in\Dt(o_i)$ for  $i = 1,2$, $\si_i\in\Dt(a_i)$  for $i=1,2$,
such that $o_1\perp a_1$, $o_2\perp a_2$. For
$T\in (\rho_1 ,\rho_2)$, $S\in (\si_1 ,\si_2 )$ we have
\[
 \ \te(n_2,m_2)\cdot T \times S \ = \
 S\times T\cdot\te(n_1,m_1).
\]
where $n_1,n_2,m_1,m_2$ denote, respectively, the multiplicities of $\rho_1,\rho_2,\si_1$ and $\si_2$.
\end{lemma}
To prove the statement we need three preliminary lemmas.
\begin{lemma}
\label{B:2}
Let $\rho\in\Dt(o)$, $\si\in\Dt(a)$ such that $o\perp a$.
If $b\in\K, \ o,a\perp b$  and $A\in\A(b)$ then we have
$  \ \te(n,m)\cdot\rho\si(A) \ = \ \si\rho(A)\cdot\te(n,m)$.
\end{lemma}
\begin{proof}
Notice that $\rho\si(A) =  1_{\rho}\otimes 1_{\si}\cdot A\otimes
1_{n_{\si}}$ and $\si\rho(A)  =  1_{\si}\otimes 1_{\rho}\cdot A\otimes
1_{n_{\rho}}$ because of localization of $\rho$ and $\si$.
As $1_\rho$ and $1_\si$ have commuting values, by (\ref{B:0})
we have the proof.
\end{proof}
\begin{lemma}
\label{B:3}
Let $\rho_1,\rho_2\in\Dt(o)$, $\si_1,\si_2\in\Dt(a)$ such that
 $o \perp a$  and let  $T\in(\rho_1 ,\rho_2), S\in(\si_1,\si_2)$.
Then we have   $\te(n_2,m_2)\cdot T\otimes S \ = \
         S\otimes T \cdot\te(n_1,m_1 )$.
\end{lemma}
\begin{proof}
$(T\times S)_{i,j} \ = \
        T_{i_{1},k}\rho_1(S_{i_{2},j_{2}})_{k,j_{1}}$ =
$T_{i_{1},j_{1}} S_{i_{2},j_{2}} \ = \ (T\otimes S')_{i,j}$.
Similarly $S\times T \ = S\otimes T$.
Since the values of $T$ and $S$ commute, the proof follows  by (\ref{B:0}).
\end{proof}
\begin{lemma}
\label{B:4}
If $\rho\in\Dt(o), \ \si\in\Dt(a)$ such that
$o \perp a$
then  $\te(n,m)\in (\rho\si,\si\rho)$.
\end{lemma}
\begin{proof}
For each $b\in\K, \ o\cup a\subseteq b$ let us take
$a_1,o_1,c,d\in\K$
such that $o_1,a_1  \perp b $,
$(o \cup o_1) \subset d$, $(a\cup a_1)\subset c$ and  $d\perp c$.
Moreover, let $U\in(\rho,\rho_1)$, $V\in(\si,\si_1)$ be unitaries and
$\rho_1\in\Dt(o_1)$, $\si_1\in\Dt(a_1)$. For $A\in\A(b)$
by Lemma \ref{B:2}
and Lemma \ref{B:3} we have
\begin{align*}
 \te(n,m)\cdot\rho\si(A) & =
  \te(n,m)\cdot(U^{+}\times V^{+})\cdot\rho_1\si_1(A)
 \cdot(U\times V)  \\
  & =  (V^{+}\times U^{+})\cdot\te(n_1,m_1)\cdot\rho_1\si_1(A)
  \cdot(U\times V)  \\
 &    =  (V^{+}\times U^{+})
  \cdot\si_1\rho_1(A)\cdot\te(n_1,m_1)\cdot
  (U\times V) \\
  & =  (V^{+}\times U^{+})
  \cdot\si_1\rho_1(A)\cdot(V\times U)\cdot
  \te(n,m)
    =   \si\rho(A)\cdot\te(n,m).
 \end{align*}
\end{proof}
\begin{proof}[Proof of Lemma  \ref{B:1}]
We use a standard deformation argument. Let $U\in(\si_3,\si_1)$ be
unitary with $\si_3\in\Dt(a_3)$ and $(a_3\cup a_1)\perp o_1$.
By Lemma \ref{B:3}
$\te(n_1,m_1)\cdot 1_{\rho_1}\times U$
= $U\times 1_{\rho_1}\cdot\te(n_1,m_3)$.
Setting  $S_1\defi S\cdot U$  we have that
$\te(n_2,m_2)\cdot T\times S_1$
= $S_1\times T\cdot\te(n_1,m_3)$ if, and only if,
$\te(n_2,m_2)\cdot T\times S$
= $S\times T\cdot\te(n_1,m_1)$.
Similarly we can move the support\footnote{By support we mean the
double cone where the object is localized.} of $\rho_1$  without changing
the statement of the lemma. If the number of spatial dimensions
is bigger than 1, by a finite number of displacements of the supports
we can reduce the problem to the trivial situation of the
Lemma \ref{B:3}.
\end{proof}
\begin{teo}
\label{B:5}
The category $\Bim$ has a unique symmetry $\eps$
satisfying
\[
\eps(\rho,\si) \ = \  \te(n,m)
\]
whenever  $\rho$ and $\si$ are localized in mutually spacelike double cones.
\end{teo}
\begin{proof} Let us observe that if  $\eps$ is a symmetry satisfying
the relation in the statement, given two unitaries
$U\in(\rho,\rho_1)$, $V\in(\si,\si_1)$ such that
$\rho_1\in\Dt(o_1)$, $\si_1\in\Dt(a_1)$ where $o_1\perp a_1$, then we have
$\te(n_1,m_1)\cdot U\times V$ = $ \eps(\rho_1,\si_1)\cdot  U\times V$
 = $ V\times U\cdot\eps(\rho,\si)$. According to this observation
we define
\begin{equation}
\label{B:6}
\eps(\rho,\si) \ \defi \ V^{+}\times U^{+}\cdot
  \te(n_1,m_1)\cdot U\times V.
\end{equation}
The definition of $\eps$ does not depend on the choice of
$U$, $V$. Indeed, given two unitaries
$U_1\in(\rho,\rho_2)$, $V_1\in(\si,\si_2)$ such that
$\rho_2\in\Dt(o_2)$, $\si_2\in\Dt(a_2)$ where $o_2\perp a_2$,
then  by Lemma \ref{B:1} we have
$\te(n_2,m_2)\cdot(U_1U^{+}\times V_1{V}^{+} =
(V_1{V}^{+} \times U_1U^{+})\cdot\te(n_1,m_1)$. Hence
\[
(V^{+}_1\times U^{+}_1)\cdot\te(n_2,m_2)\cdot(U_1\times V_1) =
 {V}^{+}\times U^{+}\cdot\te(n_1,m_1)\cdot U\times V  .
\]
We now prove that $\eps$ is a symmetry for $\Bim$.
Let $S\in(\rho,\tau)$, $T\in(\si,\be)$ and let
$W\in(\tau,\tau_1)$, $R\in(\be,\be_1)$ be unitaries such that
$\tau_1$ and $\be_1$ are localized in spacelike double cones.
By Lemma \ref{B:1} we have
$\te(l_1,k_1)\cdot(WSU^{+})\times (RTV^{+})
    =  (RTV^{+})\times(WSU^{+})\cdot\te(m_1,n_1)$,
therefore, multiplying the r.h.s. of this identity by
$R^{+}\times W^{+}$ and the l.h.s. by  $U\times V$
we obtain:
\[
\eps(\tau,\be)\cdot S\times T  =
   T\times S\cdot\eps(\rho,\si).
\]
Now, by using (\ref{B:6}) we have
\begin{align*}
\eps(\rho,\si)\cdot\eps(\si,\rho)
 & = V^{+}\times U^{+}\cdot \te(n_1,m_1)\cdot
    1_{\rho_1}\times 1_{\si_1}
    \cdot \te(m_1,n_1)\cdot V\times U \\
  & =  V^{+}\times U^{+}\cdot \te(n_1,m_1) \cdot \te(m_1,n_1)\cdot V\times U
   \ =  \ 1_{\si\rho}.
\end{align*}
Finally, let $X\in(\ga,\ga_1)$ be unitary, such that
$\ga_1\in\Dt(b_1)$ where $b_1\perp o_1$ and $(b_1\cup o_1)\perp a_1$.
Observing that
$\te(n_1,m_1)\times 1_{\ga_1}\cdot 1_{\rho_1}\times\te(l_1,m_1) =
        \te(n_1l_1,m_1)$  we have
\begin{align*}
\eps(\rho,\si)\times & 1_{\ga}\cdot
   1_{\rho}\times\eps(\ga,\si)  = \\
&=
 (V^{+}\times U^{+}\times 1_{\ga})\cdot\te(n_1,m_1)\times
  1_{\ga}\cdot(U\times V\times 1_{\ga})\cdot \\
 & \qquad\qquad  \cdot(1_{\rho}\times V^{+}\times X^{+})\cdot
 1_{\rho}\times\te(l_1,m_1)\cdot
 (1_{\rho}\times X\times V)   \\
&=  (V^{+}\times U^{+}\times X^{+})\cdot\te(n_1,m_1)\times
 1_{\ga_1}\cdot 1_{\rho_1}\times\te(l_1,m_1)\cdot
 (U\times X\times V)\\
& =  (V^{+}\times U^{+}\times X^{+})\cdot \te(n_1l_1,m_1)
      \cdot (U\times X\times V)
 = \eps(\rho\ga,\si).
\end{align*}
This  completes  the proof.
\end{proof}
\subsection{Net-left inverses, presheaf-left inverses and homogeneity}
\label{C}
The net-left inverse of an object of $\Bim$ is the obvious generalization
of the concept of left inverse of unital endomorphisms
of a $\mathrm{C}^*$-algebra  to the case where $\rho$ is a
morphism of the net.
\begin{df}
\label{Ca:1}
A \textit{net-left inverse} $\varphi$ of an object $\rho\in\Dt$ is
a nonzero completely positive normalized linear map
$\varphi: (\A\otimes\mathbb{M}_{n_\rho})_{\rho(1)}\longrightarrow \A$
fulfilling the relation
\[
\varphi(B\cdot \rho(A)) \ = \ \varphi(B) \cdot A
\]
for each $A\in\A$, $B\in (\A\otimes\mathbb{M}_{n_\rho})_{\rho(1)}$.
\end{df}
Clearly  the faithfulness of the objects
is  necessary  for the existence of net-left
inverses. It is less obvious to see that it is also  sufficient.
The physical idea used in DHR analysis  to show the existence
 of left inverses for unital endomorphisms,
 which is based on a charge transfer chain to infinite,
 can be suitably adapted to our case. Unfortunately it does not work.
 The reason is clear:  as the objects are nonunital morphisms
 it is not possible to check whether the chain has a trivial limit or not.
However, there is  another idea that can be used
to prove the existence of net-left inverses.
Notice that
a net-left inverse of $\rho$ is also a linear map
extending $\rho^{\sst{-1}}$ to the codomain algebra
$(\A\otimes\mathbb{M}_{n_\rho})_{\rho(1)}$ of $\rho$.
Such an extension can be obtained by generalizing, to our case,
an argument used for  unital endomorphisms~\cite{Fre}.
\begin{prop}
\label{Ca:2}
Each faithful object has a net-left inverse.
\end{prop}
\begin{proof}
Let $\rho\in\Dt$ be faithful and let $\Omega\in\mathcal{H}_o$.
Then we can define a state $\omega$ as
$\omega(A)\defi (\Omega,\rho^{\sst{-1}}(A)\Omega)$
          for $A\in \rho(\A)$.
Since the inclusion
$\rho(\A)\subseteq (\A\otimes\mathbb{M}_{n_\rho})_{\rho(1)}$
preserves the identity, there is a state $\omega'$
of the algebra $(\A\otimes\mathbb{M}_{n_\rho})_{\rho(1)}$ which extends
$\omega$.
Let $(\mathcal{H}', \pi',\Omega')$ be the GNS construction
associated with $\omega'$ and let us define
$V A \Omega \defi
        \pi'(\rho(A))\Omega'$ for  $A\in\A.$
As $\A$ is irreducible  and $\omega'$ is an extension
of $\omega$, $V:\mathcal{H}_o\longrightarrow
                    \mathcal{H}'$ is an isometry
fulfilling the relation $V A = \pi'(\rho(A))V$ for $A\in\A$.
Now, by setting $\varphi(A)\defi V^*\pi'(A)V$ for
$A\in (\A\otimes\mathbb{M}_{n_\rho})_{\rho(1)}$,
one easily checks that $\varphi$ is a net-left inverse of $\rho$.
\end{proof}
A left inverse (see the definition in appendix) is uniquely
associated with a net-left inverse of an object $\rho$.
To show this  we will need to
represent an element $E\in(\rho\si,\rho\tau)$
as a  $n_\si\times n_\tau$ matrix  with values
$[E]_{i,j}$ in $(\A\otimes\mathbb{M}_{n_\rho})_{\rho(1)}$
(see appendix for more details).
\begin{prop}
\label{Ca:3}
Let $\varphi$ be a net-left inverse of an object $\rho$. Then there exists
a unique positive normalized left inverse $\Phi$ of $\rho$ verifying
for each $\si,\tau\in\Dt$  the relation
\begin{equation}
\label{Ca:4}
\Phi_{\si,\tau}(E)_{i,j} \ = \ \varphi([E]_{i,j})\qquad
  E\in(\rho\si,\rho\tau)
\end{equation}
for $i = 1,\ldots n_\tau$, $ j=1,\ldots, n_\si$.
\end{prop}
\begin{proof}
The proof of the uniqueness follows once we have shown that
the relation (\ref{Ca:4}) defines a left inverse of $\rho$.
So, let $\Phi$ be the set of the linear maps $\Phi_{\si,\tau}$
for $\si,\tau\in\Dt$ defined by (\ref{Ca:4}).
$\Phi$ is obviously normalized, and it is positive because $\varphi$
is completely positive. In the rest of the proof the relations
(\ref{Xb:4}),(\ref{Xb:5}) and (\ref{Xb:6}) will be used.
Let $A\in\A$ and $E\in(\rho\si,\rho\tau)$, then
\begin{multline*}
(\Phi_{\si,\tau}(E)\cdot \si(A))_{i,j}  =
          \varphi([E]_{i,k}) \cdot \si(A)_{k,j}
        = \varphi([E]_{i,k}\cdot \rho(\si(A)_{k,j}))\\  =
           \varphi([ \ E\cdot \rho(\si(A)) \ ]_{i,j})
         =  \varphi([ \ \rho(\tau(A))\cdot E  \ ]_{i,j}) =
          (\tau(A)\cdot \Phi_{\si,\tau}(E))_{i,j}.
\end{multline*}
Hence $\Phi_{\si,\tau}(E)\in(\si,\tau)$. Given $S\in(\al,\si)$,
$T\in(\be,\tau)$, then we have
\begin{multline*}
\Phi_{\al,\be}(1_\rho\times T^+ \cdot E\cdot 1_\rho \times S)_{i,j}  =
   \varphi([1_\rho\times T^+ \cdot E\cdot 1_\rho \times
   S]_{i,j}) \\
   = \varphi( \rho(T^+_{i,k}) \cdot [E]_{k,l} \cdot\rho(S_{l,j})) =
    T^+_{i,k}\cdot \varphi([E]_{k,l}) \cdot S_{l,j} =
     (T^+\cdot\Phi_{\si,\tau}(E)\cdot S)_{l,j}.
\end{multline*}
In a similar way one can show that
$\Phi_{\si\pi,\tau\pi}(X\times 1_\pi) = \Phi_{\si,\tau}(X)\times
1_\pi$ for $X\in(\rho\si,\rho\tau)$. This completes the proof.
\end{proof}
Summing up, faithfulness is a necessary and sufficient condition
for the existence  of net-left inverses.
Moreover, for  each   net-left
inverse there is an associated left inverse of the object. It is worth
observing that this excludes neither  the existence
of nonfaithful objects with left inverses nor  the existence
of objects without left inverses. \\[5pt]
\indent We now turn to the presheaf-left inverse.
\begin{df}
\label{Cb:2}
A \textit{presheaf-left inverse of an object $\rho$  in $\Dt(o)$}
is defined as  a  collection
$\varphiB \defi
 \{ \ _a\varphi: (\A(a)'\otimes\mathbb{M}_{n_\rho})_{\rho(1)} \
\longrightarrow \ \A(a)'\ | \   a\in\K, \ a\perp o   \}$
of nonzero completely positive normalized linear maps verifying
the  relations: \\[3pt]
i)  \  $ _a\varphi\upharpoonright
        (\A(b)' \otimes\mathbb{M}_{n_\rho})_{\rho(1)}
         =  {_b\varphi}$,   \qquad $  \ \forall b\in\K, \  b\perp o,
     \  a\subseteq b $\\[3pt]
ii) \  $   _a\varphi(B \cdot {_a\rho}(A)) \ = \
           _a\varphi(B)\cdot A,$
   $\qquad \ \ \forall A\in\A(a)', \
    B\in(\A(a)'\otimes\mathbb{M}_{n_\rho})_{\rho(1)}$.
\end{df}
 It is worth observing that the definition of  presheaf-left inverse
   depends on the double cone $o$ where the object $\rho$ is
   localized. This is so because
   the inclusion $_a\rho(\A(a)')\subset
   (\A(a)' \otimes\mathbb{M}_{n_\rho})_{\rho(1)}$  is verified only for
   double cones $a$ which are spacelike separated from~$o$. Morever,
   notice that by (\ref{Aa:3}) and (\ref{Aa:4}),  a
  presheaf-left inverse of  $\rho\in\Dt(o)$ can be seen as
  a linear map from the codomain of $\rhoB\in\Dt^{\sst{\perp}}(o)$
   onto  its domain
  which is, by relation i),
  compatible with the presheaf structure and
 fulfills relation ii).\\[5pt]
\indent Now, if $\rho\in\Dt(o)$ has a presheaf-left inverse then
each $\si\in\Dt(b)$ equivalent to $\rho$, with $o\subseteq b$,
has a presheaf-left inverse. In fact,  given a unitary $U\in(\rho,\si)$,
the collection of  linear maps defined
as  $_a\varphi(U^+ B  U)$ for
$B\in(\A(a)'\otimes\mathbb{M}_{n_\si})_{\si(1)}$ and for each
$a\perp b$ is a presheaf-left inverse of $\si$.
This argument cannot be applied
to  the elements of $[\rho]$ localized in double cones
which do not contain $o$. Hence, in general, having a presheaf-left inverse
is not a property of the equivalence class of the object.
This leads to the following
\begin{df}
\label{Cb:2a}
We say that an object $\rho$ of $\Bim$ is \textit{homogeneous} if each
element of its equivalence class has presheaf-left inverses, namely
if for each $a\in\K$,  any $\si\in[\rho]$ localized in $a$
has, as an  element of $\Dt(a)$, a presheaf-left inverse.
\end{df}
Concerning homogeneity and existence of presheaf-left inverses, in this
section   we will limit ourselves  to the following remark.
If $\rho\in\Dt(o)$ has a presheaf-left
inverse, then $_a\rho$ is a  faithful morphism  for each $a\perp o$.
This does  not imply the double faithfulness of
$\rho$. Double faithfulness occurs when $\rho$ is homogeneous.
Conversely, it is not clear whether in general  double faithfulness
it is enough  both for the existence of presheaf-left
inverses and for homogeneity.
We only know that this happens in the particular case
of doubly faithful simple objects (see next section).\\[5pt]
\indent As a  consequence of the definition of presheaf-left inverse we have
the following
\begin{prop}
\label{Cb:3}
Let $\varphiB$ be a presheaf-left inverse of $\rho\in\Dt(o)$. Then
there exists a unique net-left inverse $\varphi$ of $\rho$ such that
for each $a,b\in\K$ and $b\perp a,o$  we have
\begin{equation}
\label{Cb:4}
\qquad \varphi(A) \ = \ _b\varphi(A)
    \qquad A\in(\A(a)\otimes\mathbb{M}_{n_\rho})_{\rho(1)} .
\end{equation}
\end{prop}
\begin{proof}
Since  $a^{\sst{\perp}}\cap o^{\sst{\perp}}$  is pathwise connected,
the compatibility of $\varphiB$ implies (in the same
way as in Proposition \ref{Aa:2}.a)  that
the relation (\ref{Cb:4}) defines a net-left inverse of $\rho$.
\end{proof}
The relations (\ref{Cb:4}), (\ref{Ca:4})
yield  a correspondence between presheaf-left inverses and
left inverses. Namely, given
a presheaf-left inverse $\varphiB$ of $\rho\in\Dt(o)$ we have
\begin{equation}
\label{Cb:5}
\{ \mbox{\textit{presheaf-left  inverses  of }} \rho\in\Dt(o)  \}   \ni
\varphiB \ \ \xrightarrow{(\ref{Cb:4})}  \ \ \varphi \ \
           \xrightarrow{(\ref{Ca:4})}  \ \ \Phi\in
   \{ \mbox{\textit{left  inverses  of }}  \rho \}
\end{equation}
We denote by $l(\varphiB)$ the left inverse defined by relation
(\ref{Cb:5}) and call it the left inverse \textit{associated} with $\varphiB$.
Notice that  for each  $E\in(\rho,\rho)$ we have
\[
l(\varphiB)_{\io,\io}(E) \ = \ _a\varphi(E)
\qquad \forall a\subset o^{\sst{\perp}}  \ \ \Longrightarrow
\ \ _a\varphi(E)= c\cdot 1 \qquad
\forall a\in\K, \ a\perp o
\]
because of the irreducibility of $\io$. \\
\indent We conclude this section by studying how the correspondence
$\varphiB\longrightarrow l(\varphiB)$
behaves under the categorical operations. Let $\varphiB$,$\varphiB_1$ and
$\varphiB_2$ be presheaf-left inverses of
$\rho,\rho_1,\rho_2\in\Dt(o)$
respectively.
Given two isometries $W_i\in(\al,\rho_i) \ $ for $i=1,2$ verifying
$W_1 W^{+}_1 + W_2 W^{+}_2 = 1_{\alpha}$, and given an isometry
$V\in(\be,\rho)$ ($VV^+ \defi E$ ), 
the  linear maps
$\varphiB_{1}\circ\varphiB_{2}$,
$\varphiB_{1}\oplus^s\varphiB_{2}$ for $s\in[0,1]$ and
$\varphiB^{\sst{E}}$ defined as
\begin{align}
\label{Cb:6a}
 &  _a(\varphi_{1}\circ\varphi_{2})(A)  \defi
 {_a\varphi_{1}}(_a\varphi_{2}([A]))  &&\hspace{-0.8cm}
A\in (\A(a)'\otimes\mathbb{M}_{n_2 n_1})_{\rho_2\rho_1(1)} \\
\label{Cb:6b}
&  _a\left(\varphi_{1}\oplus^s\varphi_{2}\right)\hspace{-0.1cm}(B)  \defi
 s \  _a\varphi_{1}(W^+_1  B W_1)
        +
  (1\hspace{-0.1cm}-\hspace{-0.1cm}s) \ _a\varphi_{2}(W^+_2  B  W_2)
   && B\in (\A(a)'\otimes\mathbb{M}_{n_\alpha})_{\alpha(1)}\\
\label{Cb:6c}
&  _a\varphi^{\sst{E}}(C)  \defi  {_a\varphi(E)}^{-1}
  {_a\varphi}(V  C   V^+) \qquad \mbox{if }
 {_a\varphi(E)}\ne 0
  &&  C\in (\A(a)'\otimes\mathbb{M}_{n_\be})_{\be(1)}
\end{align}
for each $a \perp o$, are,  respectively, presheaf-left inverses of
$\rho_2\rho_1$, $\al$ and $\be$ as elements of  $\Dt(o)$.
The definitions (\ref{Cb:6a}) and (\ref{Cb:6b}) entail
that the existence of presheaf-left inverses
and the homogeneity are stable properties under tensor products and
direct sums. This cannot be asserted for subobjects because
$\varphiB^{\sst{E}}$ exists only if the scalar
${_a\varphi(E)}\ne 0$. Now, note that the same constructions we have made
for presheaf-left inverses  can be made for the associated left inverses
$l(\varphiB)$, $l(\varphiB_1)$ and $l(\varphiB_2)$
(see (\ref{Xa:2a}), (\ref{Xa:2b}) and
(\ref{Xa:2c})).
One can easily show  that the correspondence  $\varphiB
\longrightarrow l(\varphiB)$ is compatible: for each
$\si,\tau\in\Dt$ we have
\begin{align}
\label{Cb:7a}
 \left(l(\varphiB_{1})\circ l(\varphiB_{2})\right)_{\si,\tau}
   & = l(\varphiB_{1} \circ \varphiB_{2})_{\si,\tau} && \\
\label{Cb:7b}
  \left(l(\varphiB_{1}) \oplus^s l(\varphiB_{2})\right)_{\si,\tau} & =
l(\varphiB_{1}\oplus^s\varphiB_{2})_{\si,\tau}  &&
      \forall s\in[0,1]\\
\label{Cb:7c}
  l(\varphiB)^{\sst{E}}_{\si,\tau} & =  l(\varphiB^{\sst{E}})_{\si,\tau} &&
    \mbox{ if } l(\varphiB)_{\io,\io}(E) \ne 0 \
\end{align}
These relations allow us to work with the more tractable associated
left inverses  rather than with the presheaf-left inverses. In
particular, by (\ref{Cb:7c}) the existence of the
presheaf-left  inverse $\varphiB^{\sst{E}}$ for the subobject $\be\in\Dt(o)$
is equivalent to the condition $l(\varphiB)_{\io,\io}(E)\ne 0$.
\subsection{Simple objects}
\label{D}
In this section we study the simple objects of $\Bim$, namely
objects characterized by the following property: $\ga\in\Dt$ is
\textit{simple} if $\eps(\ga,\ga) \ = \chi_\ga \cdot 1_{\ga^2}$
where $\chi_\ga \in\{1,-1\} $.
In this section we show that each doubly faithful simple object
is homogeneous. \\[5pt]
\indent Let us start by noting that if
$\ga\in\Dt$ is  simple then each element of $[\ga]$ is simple.
Now, if $\ga$ has a left inverse $\Phi$ then the following properties
are equivalent
\[
 \Phi_{\ga,\ga}(\eps(\ga,\ga)) = \pm 1_\ga \ \iff \
  \ga \mbox{ is simple }  \ \iff \
  (\ga^{\sst{2}},\ga^{\sst{2}}) = \mathbb{C}\cdot 1_{\ga^2}
\]
(see Proposition \ref{Xa:5}).
Moreover the l.h.s. and the r.h.s. relations imply that $\ga$ is
irreducible. Hence when a simple object $\ga$ has a left inverse
we can say that $[\ga]$  is a \textit{simple sector}. \\
\indent Progress now results from  studying the structure of the faithful
simple objects.
\begin{lemma}
\label{D:1}
Let $\ga\in\Dt(o)$ be a faithful simple object and
 let $\varphi$ be a net-left
inverse of $\ga$. For each $b\in\K, \ o\perp b$ and for each
unitary $U\in(\si,\ga)$ such that
$\si$ is localized in $b$ we have
\[
  \si(\varphi(B)) \ = \
      U^+ \cdot B \cdot U
  \qquad B\in (\A(o)\otimes\mathbb{M}_{n})_{\ga(1)}.
\]
In particular $\varphi\upharpoonright
(\A(a)\otimes\mathbb{M}_{n})_{\ga(1)}$ is injective
for each $a\in\K, \ o\subseteq a$.
\end{lemma}
\begin{proof}
In the proof the relations
(\ref{Xb:4}),(\ref{Xb:5}) and (\ref{Xb:6}) will be used.
Observing that $1_\si\otimes B  =  \si(B)$, because $o\perp b$, we have
\begin{align*}
[1_\si\otimes   B]_{i,j}  &=  \si(B_{i,j})
 =   U^+\cdot \ga(B_{i,j}) \cdot U \\
& =   U^+ \cdot \ga((1_\ga)_{i,t}) \cdot
       \ga(B_{t,s}) \cdot
       \ga( (1_\ga)_{s,j}) \cdot U
 = [U^+\times 1_\ga \cdot \ga(B) \cdot U\times 1_\ga]_{i,j}
\end{align*}
By using this identity we have
\begin{align*}
B& \otimes  1_\si  = \\
  & = \theta(n_\si,n_\ga) \cdot 1_\si\otimes B \cdot
                     \theta(n_\ga,n_\si)
   =    \theta(n_\si,n_\ga) \cdot
        U^+\times 1_\ga \cdot \ga(B) \cdot U\times 1_\ga \cdot
      \theta(n_\ga,n_\si) \\
  & =  \eps(\si,\ga)\cdot
      U^+\times 1_\ga \cdot \ga(B) \cdot U\times 1_\ga \cdot
      \eps(\ga,\si)
   =  1_\ga\times U^+ \cdot \eps(\ga,\ga) \cdot \ga(B) \cdot
       \eps(\ga,\ga) \cdot 1_\ga \times U \\
  & =  1_\ga\times U^+ \cdot \ga(B) \cdot 1_\ga \times U
\end{align*}
In conclusion we have
\begin{align*}
\si(\varphi(B))_{i,j} &
   =\varphi(B)   \cdot \si(1)_{i,j}  = \varphi(B\ga(\si(1)_{i,j})) =
    \varphi([B\otimes 1_\si]_{i,j}) \\
& = \varphi([1_\ga\times U^+]_{i,t} \cdot [\ga(B)]_{t,s} \cdot [1_\ga \times U]_{s,j})
  =  \varphi(\ga(U^+_{i,t}) \cdot \ga(B_{t,s}) \cdot \ga(U_{s,j}))\\
& =    (U^+\cdot B\cdot U)_{i,j}
\end{align*}
\end{proof}
\begin{teo}
\label{D:2}
Let $\ga$ be a simple object. If $\ga$ is faithful, then
$\ga:\A\longrightarrow (\A\otimes\mathbb{M}_n)_{\ga(1)}$
is an isomorphism and  $\ga^{\sst{-1}}$ is the unique net-left
inverse of $\ga$.
\end{teo}
\begin{proof}
Let  $\ga$ be localized in $o$.  For each
$a\in\K, \ o\subseteq a$,  let $U\in(\si,\ga)$ be unitary such that
$\si$ is localized in a double cone spacelike  separated  from $a$.
Observing that $\varphi((\A(a)\otimes\mathbb{M}_n)_{\ga(1)}) =
\A(a)$ then by the previous lemma we have
\begin{align*}
\si(\varphi (A^+B)) =
      U^+  A^+B  U & =
       U^+  A^+ U   U^+ B  U \\
  & = \si(\varphi(A))^* \cdot \si(\varphi(B))
      = \si( \ \varphi(A)^*  \varphi(B) \ )
\end{align*}
$A,B\in(\A(a)\otimes\mathbb{M}_n)_{\ga(1)}$.
Since $\si$ is faithful,
$\varphi(A^+B) = \varphi(A)^*\varphi(B)$
that is,  $\varphi$ is a morphism. It is surjective and injective
therefore  $\varphi  \ = \ \ga^{\sst{-1}}$ on
$(\A(a)\otimes\mathbb{M}_n)_{\ga(1)}$. Now, the proof follows by
continuity of $\varphi$.
\end{proof}
\begin{cor}
\label{D:3}
Let $\ga$ be a simple object. Then
$\ga$ is doubly faithful if, and only if, $\ga$ is homogeneous.
In particular if $\ga\in\Dt(o)$ is doubly faithful,  then
$\{_a\ga : \A(a)'  \longrightarrow
(\A(a)'\otimes\mathbb{M}_{n_\ga})_{\ga(1)}, \  a \perp o \}$
is a presheaf isomorphism and
$\gaB^{\sst{-1}}\defi \{ _a\ga^{\sst{-1}},  \ a\perp o \}$ is the unique
presheaf-left inverse of $\ga\in\Dt(o)$.
\end{cor}
\begin{proof}
Let us  assume that $\ga$ is localized in $o\in\K$.  It is only
a matter of calculation  to check that, for each $a\perp o$,
$_a\ga^{\sst{-1}}(\A(a^{\sst{\perp}})\otimes\mathbb{M}_{n_\ga})_{\ga(1)} \subseteq
\A(a)'$.
Applying $_a\ga$ to this  inclusion  and observing that,
by Proposition \ref{Aa:2}.c,
$_a\ga(\A(a)')\subseteq (\A(a)'\otimes\mathbb{M}_{n_\ga})_{\ga(1)}$
we have
\[
(\A(a^{\sst{\perp}})\otimes\mathbb{M}_{n_\ga})_{\ga(1)} \ \subseteq \ _a\ga(\A(a)')
\subseteq \ (\A(a)'\otimes\mathbb{M}_{n_\ga})_{\ga(1)}.
\]
Passing to the weak closure,  by Haag duality, we have
$_a\ga(\A(a)')=(\A(a)'\otimes\mathbb{M}_{n_\ga})_{\ga(1)}$.
This implies that
$\{ _a\ga:  \A(a)'  \longrightarrow
(\A(a)'\otimes\mathbb{M}_{n_\ga})_{\ga(1)},  \ a \perp o\}$
is a presheaf isomorphism and that
$\gaB^{\sst{-1}}$, defined as above,
is the unique  presheaf-left inverse
of $\ga\in\Dt(o)$. Since double faithfulness is stable under
equivalence, each element of the equivalence class of $\ga$
admits presheaf-left inverses. Hence, $\ga$ is homogeneous; the converse
statement is contained in the observation following
Definition \ref{Cb:2a}.
\end{proof}
\subsection{The category of  objects with finite statistics}
\label{E}
The only known  way  to classify  the statistics of sectors
of a tensor $\mathrm{C}^*$-category with a symmetry, is the
one followed in DHR analysis  and based on
using  left inverses. But this procedure
might not be applicable to all the sectors of $\Bim$ because,
as observed in  Section \ref{C}, we cannot  exclude the existence
of objects without left inverses. Disregarding the sectors associated
with this kind of objects, we could proceed as in DHR analysis
and classify the statistics of the sectors as finite or infinite.
However, we will see that objects with conjugates have finite
statistics; henceforth  we will confine ourselves to this case.\\[5pt]
\indent Let $\mathrm{A_d},\mathrm{S_d}\in(\rho^{\sst{d}},\rho^{\sst{d}})$ be
the (anti)symmetrizer associated with
$\eps^d_\rho$.
\begin{df}
\label{E:1}
We say that an object of $\Bim$ has
\textit{finite statistics} if it is finite direct sum of irreducibles
$\rho$ fulfilling the following conditions:
there is a 3-tuple $(d,\ga,V)$
where $d$ is an integer, $\ga$ is a faithful simple object
and $V\in(\ga,\rho^{\sst{d}})$ is an isometry satisfying one
of the following alternatives:
\[
 B) \ \   VV^+ \ = \  \mathrm{A_d} \qquad \mbox{or}
\qquad
       F)  \ \  VV^+ \ = \  \mathrm{S_d}.
\]
We denote by $\Delta_{\mathrm{f}}$ the set of the objects with finite
statistics and by $\Bim_{\mathrm{f}}$ the full subcategory of $\Bim$
whose objects have finite statistics.
\end{df}
The finiteness of the statistics
is stable under equivalence.  Moreover,
each object $\rho\in\Delta_\mathrm{f}$ has left inverses.
By definition of $\Delta_\mathrm{f}$
and by (\ref{Xa:2b}) it is enough to prove this in the case where
$\rho$ is irreducible.
Let $(d,\ga,V)$ be the 3-tuple associated with $\rho$.
Since $\ga$ is faithful, it has a left inverse $\Phi$,
Proposition \ref{Ca:2}. Then
\begin{equation}
\label{E:2}
\Psi_{\si,\be}(X) \defi
\Phi_{\si,\be}( V^+\times 1_\be \cdot 1_{\rho^{d-1}}\times X \cdot V\times
1_\si ) \qquad X\in(\rho\si,\rho\be)
\end{equation}
defines a left inverse of $\rho$. \\
\indent Our definition of objects with finite statistics
is equivalent to the usual one, as the following propositions show.
\begin{prop}
\label{E:3}
Let $\rho$ be irreducible. Then the following assertions are equivalent: \\
a) $\rho$ has finite statistics; \\
b) each left inverse $\Psi$ of $\rho$ verifies the relation
$\Psi_{\rho,\rho}(\eps(\rho,\rho))= \la  \cdot 1_\rho \ne 0$
\end{prop}
\begin{proof}
b) $\Rightarrow$ a) follows directly from DHR analysis,
\cite[Section III]{DHR4}.
We show  a) $\Rightarrow$ b) only in the case $B)$.
In a similar way, one can easily check that this holds
in the case $F)$ as well. Furthermore,
it is possible to prove, as in DHR analysis,
that if $\Psi$ is a left inverse of
$\rho$ such that $\Psi_{\rho,\rho}(\eps(\rho,\rho))= \la  \cdot 1_\rho$
then the real number $\la$ is an invariant
of the equivalence class $[\rho]$. Hence, the proof follows
once we have shown the existence of one left inverse verifying
the relation in the statement. In order to prove this
let  $(d,\ga,V)$ be the 3-tuple associated with $\rho$,
and let $\Phi$ be a left inverse of $\ga$. Moreover,
let $\Psi$  be the left inverse of
$\rho$ defined  by using $\Phi$ in (\ref{E:2}).
$\rho$  being  irreducible then
$\Psi_{\rho,\rho}(\eps(\rho,\rho))= \la\cdot 1_\rho$.
We now prove that  $\la\ne 0$. To this aim,
we need some preliminary observations.
First, we recall the following formula \cite[Lemma 5.3]{DHR3}:
\[
\Psi^{\circ_d}_{\io,\io}(\mathrm{A_d})  =
     d!^{\sst{-1}}\cdot (1-\la)(1-2\la)\cdots(1-(d-1)\la) \qquad  (*)
\]
where $\Psi^{\circ_d}$ is the left inverse of $\rho^{\sst{d}}$
given by the $d$-fold composition of
$\Psi$. Secondly,
$\Psi^{\circ_d}_{\rho^d,\rho^d}(\eps(\rho^{\sst{d}},\rho^{\sst{d}})) =
\la^{\sst{d}} \cdot 1_{\rho^d}$ because of (\ref{Xa:3a}).
Thirdly, if $\Psi^{\circ_d}_{\io,\io}(\mathrm{A_d}) \ne 0$,
then
\[
\tilde{\Phi}_{\si,\tau}(F) \ = \ ( \
    \Psi^{\circ_d}_{\io,\io}(\mathrm{A_d}) \ )^{\sst{-1}}\cdot
       \Psi^{\circ_d}_{\si,\tau}
    (V\times 1_\tau \cdot F \cdot V^+\times 1_\si)  \qquad
        F\in(\ga\si,\ga\tau)
\]
defines a left inverse of $\ga$ such that
\[
\begin{array}{lclr}
\chi_\ga \cdot 1_\ga & = & \tilde{\Phi}_{\ga,\ga}(\eps(\ga,\ga))
   =  (\Psi^{\circ_d}_{\io,\io}(\mathrm{A_d}))^{\sst{-1}} \cdot
   \Psi^{\circ_d}_{\ga,\ga}(V^+\times 1_\ga \cdot \eps(\ga,\ga) \cdot
  V\times 1_\ga) \\
  & =  & (\Psi^{\circ_d}_{\io,\io}(\mathrm{A_d}))^{\sst{-1}}\cdot
   \ V  \Psi^{\circ_d}_{\rho^d,\rho^d}
     (\eps(\rho^{\sst{d}},\rho^{\sst{d}})) V^+ \ = \
  (\Psi^{\circ_d}_{\io,\io}(\mathrm{A_d}))^{\sst{-1}}\cdot \la^{\sst{d}}\cdot
   1_{\ga} & (**)
\end{array}
\]
because $\eps(\ga,\ga)= \chi_\ga \cdot 1_{\ga^2}$. Now the proof that $\la\ne 0$
proceeds as follows.
If $\la$ were equal to $0$, then $\tilde{\Phi}$ would be well defined
because, by $(*)$,  $\Psi^{\circ_d}_{\io,\io}(\mathrm{A_d}) \ = \
d!^{\sst{-1}}$. This leads
to a contradiction because, by $(**)$, we should have
$\chi_\ga\cdot  1_\ga \ = \ 0$.
\end{proof}
\begin{prop}
\label{E:4}
The following assertions hold: \\
a) $\Bim_\mathrm{f}$ is closed under tensor products, direct sums and
subobjects;\\
b) $\rho\in\Delta_\mathrm{f}$ $\iff$  has a standard left inverse $\Phi$, that is
$\Phi_{\rho,\rho}(\eps(\rho,\rho))^2 = c \cdot 1_{\rho}>0$.
\end{prop}
\begin{proof}
a) The closedness under direct sums and subobjects is obvious by
   definition of $\Bim_\mathrm{f}$.
   Once we have shown that  given two  irreducibles  $\rho,\si$  with
   finite statistics  each subobject of $\rho\si$ has left inverses,
   the closedness under tensor products
   follows as in DHR analysis.
   For this
   purpose, let $\Phi,\Psi$ be  two left
   inverses of $\rho$ and $\si$  respectively.
   By Proposition \ref{Xa:3b} the left inverse $\Psi\circ \Phi$
   of $\rho\si$ is faithful.
   Hence each subobject of $\rho\si$ has a left inverse
   defined by (\ref{Xa:2c}).
b) $(\Rightarrow)$ follows as in the DHR analysis.
   $(\Leftarrow)$ By Proposition \ref{Xa:3} any standard left inverse
is faithful. Therefore  each subobject of $\rho$ has left inverses.
The rest of the proof  follows as in DHR analysis.
\end{proof}
\subsection{The selection of the relevant subcategory}
\label{Eb}
In the previous section, in order to exclude
objects without conjugates,
we introduced  the category $\Bim_\mathrm{f}$. With the same motivation,
we can affirm that this is only a preliminary
step  since  properties like
(double) faithfulness and homogeneity might fail to
hold  in  $\Bim_\mathrm{f}$.
Observing that homogeneity  implies the
other two properties, we show in this section how to
select the maximal full subcategory of $\Bim_\mathrm{f}$ with
homogeneous objects, closed under
tensor products,
direct sums and subobjects.
We claim here but will prove in the next section
that, this is the relevant subcategory. \\[5pt]
\indent To understand  the problem we are facing,
we recall that homogeneity might not be
stable  under subobjects
(see  the  observation related to (\ref{Cb:6c})).
Consequently the category we are looking for
does not correspond, in general, to the subcategory
of $\Bim_\mathrm{f}$ whose objects are homogeneous. However,
this category can be selected by  adding further conditions.
\begin{df}
\label{Eb:0}
We denote by $\Delta_\mathrm{fh}$ the subset of $\Delta_t$
whose objects are finite direct sums of irreducibles $\rho$
fulfilling the following conditions:
\begin{itemize}
\item[a)]  $\rho$ has finite statistics;
\item[b)] given the  3-tuple $(d,\ga,V)$
associated with $\rho$,
the simple object $\ga$  is doubly faithful
(or, equivalently,  homogeneous).
\end{itemize}
We denote by $\Bim_\mathrm{fh}$ the full subcategory of
$\Bim_\mathrm{f}$ whose objects
belong to $\Delta_\mathrm{fh}$.
\end{df}
Notice that the property of belonging to
$\Delta_\mathrm{fh}$ is stable under equivalence. Now,
the next proposition shows a useful characterization of the
irreducible elements of $\Delta_\mathrm{fh}$, while the subsequent
one  proves the main claim of this section.
\begin{prop}
\label{Eb:0a}
Let $\rho$ be irreducible. Then the following assertions are equivalent:\\
a) $\rho\in\Delta_\mathrm{fh}$; \\
b) $\rho$ is homogeneous and belongs to $\Delta_\mathrm{f}$.
\end{prop}
\begin{proof}
 a) $\Rightarrow$ b)  Given  $(d,\ga,V)$ be the 3-tuple associated with
$\rho$, where $\ga$ is homogeneous.
Let us assume that $\rho$ and $\ga$ are localized in the same
double cone $o$. Let $\gaB^{\sst{-1}}$  be
the presheaf-left inverse of $\ga\in\Dt(o)$ defined by Corollary
\ref{D:3}. Setting
\begin{equation}
\label{Eb:1}
_a\varphi(A) \ \defi \ _a\ga^{\sst{-1}}(V^+ \ _a\rho^{\sst{d-1}}(A)\ V)
  \qquad A\in(\A(a)'\otimes\mathbb{M}_{n_\ga})_{\rho(1)}
\end{equation}
for each $a\in\K, \ a\perp o$, we have that  the set
$\varphiB= \{{_a\varphi}, \  a\perp o \}$ is a
presheaf-left inverse of $\rho\in\Dt(o)$. Since each element
of $[\rho]$ belong to $\Delta_\mathrm{fh}$, $\rho$ is homogeneous.
$a)\Leftarrow b)$ Let $(d,\ga,V)$  be  the 3-tuple associated with
$\rho$, where in this case $\ga$ is faithful.
The proof follows once we have shown that $\ga$ is homogeneous. Let
us assume that $\rho$ and $\ga$  are localized in the same
double cone $o$ and let $\varphiB$ be a presheaf-left inverse of
$\rho\in\Dt(o)$. By Proposition \ref{E:3}
the left inverse $l(\varphiB)$ of $\rho$ associated with $\varphiB$
satisfies the relation
$l(\varphiB)_{\rho,\rho}(\eps(\rho,\rho))= \la\cdot 1_\rho \ne 0$.
Combining this with Proposition \ref{Xa:3b} we obtain
that $l(\varphiB)^{\circ_d}$ is a faithful left inverse of $\rho^{\sst{d}}$.
We now notice that  $l(\varphiB)^{\circ_d}= l(\varphiB^{\circ_d})$ because of
(\ref{Cb:7a}), where $\varphiB^{\circ_d}$ is the presheaf-left inverse of
$\rho^{_{d}}\in\Dt(o)$ given by the d-fold composition of $\varphiB$.
This entails that $l(\varphiB^{\circ_d})_{\io,\io}(VV^+)\ne 0$ and,
by (\ref{Cb:6c}), that
$\ga\in\Dt(o)$ has presheaf-left inverses. Since, by  transportability,
this argument
can be applied to each element of $[\ga]$,
$\ga$ is homogeneous.
\end{proof}
\begin{prop}
\label{Eb:2}
The following assertions hold: \\
a) $\Bim_\mathrm{fh}$  is  closed under tensor products, direct sums
and subobjects, and
its objects are homogeneous;\\
b)  $\Bim_\mathrm{fh}$ is the maximal full subcategory of $\Bim_\mathrm{f}$
closed under subobjects  and whose objects are homogeneous.
\end{prop}
\begin{proof}
a) The closedness under direct sums and subobjects is obvious.
Let $\rho_1,\rho_2\in\Delta_\mathrm{fh}$ be two irreducibles
localized in the same
double cone $o$. Since  $\rho_1,\rho_2$ are homogeneous,
both the direct sum and the tensor product of these two objects
are homogeneous
(see observation related to (\ref{Cb:6a}), (\ref{Cb:6b})).
It remains  to be proved that each subobject of
$\rho_1\rho_2$ is homogeneous.
Since $\rho_1$ and $\rho_2$ are homogeneous objects with finite statistics,
the proof follows by the same argument
used in the proof of the implication a) $\Leftarrow$ b) of the
previous proposition.
b) Let $\mathcal{C}$ be a category fulfilling the
properties written in the statement. Since
each object of $\mathcal{C}$ is a finite direct sum
of homogenous irreducible objects with finite statistics,
the proof follows from  Proposition \ref{Eb:0a}.
\end{proof}
\section{Conjugation}
\label{F}
This section concludes the investigation of Sections \ref{A}
and  \ref{AB}. We start by recalling the definition of the conjugate
of an object.
An object $\rho$  has conjugates if there exists an
object $\overline{\rho}$ and a
pair  of arrows $R\in(\io,\overline{\rho}\rho)$,
$\overline{R}\in(\io,\rho\overline{\rho})$
satisfying the \textit{conjugate equations}:
\[
\overline{R}^+ \times 1_\rho \cdot 1_\rho\times R \ = \ 1_\rho \ \ , \ \
   R^+ \times 1_{\overline{\rho}} \cdot 1_{\overline{\rho}}\times \overline{R} \
    = \ 1_{\overline{\rho}}
\]
Conjugation is a property which is stable under equivalence,
tensor product, direct sums and subobjects \cite{LR}.\\[5pt]
\indent The next result proves the assertions we have been claiming
throughout this paper:
\begin{teo}
\label{F:1}
Each object of $\Bim$ with conjugates belongs to $\Bim_\mathrm{fh}$.
\end{teo}
\begin{proof}
Given  $\rho$, $\overline{\rho}$   localized in $o$, let
$R,\overline{R}$ be a pair of arrows solving the
conjugate equations for $\rho$ and
$\overline{\rho}$. Then by setting
\[
 \Phi_{\si,\tau}(X) \ \defi \  (R^+R)^{\sst{-1}} \cdot \left(R^+\times 1_\tau \cdot
    1_{\overline{\rho}}\times X \cdot  R\times 1_\si\right) \qquad
  X\in(\rho\si,\rho\tau)
\]
for each $\si,\tau\in\Dt$, we get a left inverse $\Phi$ of $\rho$.
Since it is always possible to choose $\overline{R},R$ in a way that
$\Phi$ is standard \cite{LR}, by Proposition \ref{E:4}.b
$\rho$ has finite statistics. Now, the set of linear maps defined  as
\[
_a\varphi(A) \ \defi \ (R^+R)^{\sst{-1}} \cdot \left(R^+ \cdot {_a\overline{\rho}}(A) \cdot R\right) \qquad
  A\in(\A(a)'\otimes\mathbb{M}_{n_\rho})_{\rho(1)}
\]
for each $a\perp o$, is a presheaf-left inverse of
$\rho\in\Dt(o)$.  As  conjugation is stable under equivalence,
each element of $[\rho]$ has presheaf-left inverses. Thus $\rho$
is homogeneous. Finally,  since conjugation is stable under subobjects,
Proposition \ref{Eb:2}.b  completes the proof.
\end{proof}
Theorem \ref{F:1} states that all the relevant objects of the theory
belong to $\Bim_\mathrm{fh}$.
We claim  that $\Bim_\mathrm{fh}$  \textit{is the relevant
subcategory} of the theory. In fact,
as we will prove later,  each object of
$\Bim_\mathrm{fh}$  has conjugates  under the assumption that the
local algebras are properly infinite, and there are several reasons
for  considering this assumption as an
essential property of the reference representation:\\[5pt]
\indent \textit{First}, in  an arbitrary
globally hyperbolic spacetime the algebras of local observables
of a multiplet of $n$ Klein-Gordon fields in any Fock representation,
acted on by $U(n)$ as a global gauge
group, fulfill this property
(this result is proved in \cite{Ru} and will  be described in
a forthcoming article). \\[5pt]
\indent \textit{Secondly}, it turns out to be a necessary condition in the
following particular situation
\begin{teo}
\label{F:3}
Let $\rho,\overline{\rho}$ be two conjugate  objects
with multiplicity equal to one.
If $\rho$ is not a simple object, then
$\boldsymbol{\A}$  is a net of properly infinite von Neumann algebras.
\end{teo}
The proof is based on the following
\begin{lemma}
\label{F:4}
Let $\si,\overline{\si}$ be two conjugate objects   localized, respectively,
in $o_1,o_2\in\K$.
If $\si$, $\overline{\si}$ have multiplicity equal to one, then
each nonzero orthogonal projection $E\in(\si,\si)$
is equivalent to $1$ on  $\A(a)$ for each
$a\in\K, \ o_1\cup o_2\subseteq a$.
\end{lemma}
\begin{proof}
We recall that an object with multiplicity equal to one
is an endomorphism, in general not unital, of the algebra $\A$.
Let $V\in(\tau,\si)$ be an isometry such that $V\cdot V^+ =  E$.
The subobject $\tau$ is an endomorphism of $\A$ localized in $o_1$,
and $\tau(1) =  E$.
Given a  pair of arrows $R,\overline{R}$  solving the
conjugation equations for $\si$ and $\overline{\si}$, let us define
$S\equiv V^+\times 1_{\overline{\si}}
\cdot\overline{R}\in(\io,\tau\overline{\si})$.
By \cite[Lemma 2.1]{LR}  $S\ne 0$, hence $S' \equiv S \cdot t^{\sst{-1/2}}$
is  an isometry, where $S^+\cdot S = t\cdot 1$.
Then $1 \sim \ S'\cdot{S'}^{+}$ $\leq \tau\overline{\si}(1)$
$\leq\tau(1) \leq 1$.
Hence $E\sim 1$ on $\A(a)$.
\end{proof}
\begin{proof}[Proof of Theorem \ref{F:3}]
Let  us  assume that $\rho$ and $\overline{\rho}$ are,  respectively,
localized in $o_1$, $o_2$, and let $a\in\K, \
o_1\cup o_2\subseteq a$. By \cite[Proposition 3]{DM} it is enough
to show that
$\A(a)$ is properly infinite. Notice that $\rho^{\sst{2}}$ is reducible
because $\rho$ is not simple, Proposition \ref{Xa:5}. Hence there is
a nonzero orthogonal projection $E\in(\rho^{\sst{2}},\rho^{\sst{2}})$
such that $E\ne \rho^{\sst{2}}(1)$.
By Lemma \ref{F:4} $\rho^{\sst{2}}(1) \sim 1$
$\sim (\rho^{\sst{2}}(1) - E)$  $\sim  E$ on $\A(a)$.
If $\mathrm{Tr}$ were a finite normal trace of $\A(a)$, then the previous
relation should entail the equality
$\mathrm{Tr}(1)$ $=\mathrm{Tr}(\rho^{\sst{2}}(1))$ $= \mathrm{Tr}(E)$
$=\mathrm{Tr}(\rho^{\sst{2}}(1))-\mathrm{Tr}(E)$,
which is possible if, and only if, $\mathrm{Tr}(1)=0$.
\end{proof}
In spite of this theorem, we cannot assert  that  a general
requirement  on the reference representation for the existence
of nontrivial conjugate objects is that the local algebras are
properly infinite.
A counterexample of this assertion can be found in \cite{SV}.
The results obtained in that paper,
however, do  not affect the hypothesis
of  considering   the local algebras $\A(a)$
properly infinite as an essential property of the reference
representation because we are interested to applications
deriving from models of quantum fields theory, while that paper concerns
quantum statistical mechanics and the local algebras are defined on a
lattice.
\begin{teo}
\label{F:2}
If $\boldsymbol{\A}$ is a
net of properly infinite algebras, then
$\Bim_\mathrm{fh}$ is the full subcategory of $\Bim$
whose objects have conjugates.
\end{teo}
\begin{proof}
By virtue of the Theorem \ref{F:1}, we only have to prove
that each object of $\Bim_\mathrm{fh}$ has conjugates.
The existence  of conjugates in $\Bim_\mathrm{fh}$
is equivalent to the existence
of conjugates for its simple objects. In order to prove this, let us
consider an irreducible object $\rho$ of $\Bim_\mathrm{fh}$ and  let
$(d,\ga,V)$ be the 3-tuple associated with $\rho$. If $\ga$ has a
conjugate  $\overline{\ga}$ and $T\in(\io,\overline{\ga}\ga)$,
$\overline{T}\in(\io,\ga\overline{\ga})$  solve
the conjugate equations for $\ga$ and $\overline{\ga}$,  then  by setting
\[
\overline{\rho} \defi \overline{\ga}\rho^{\sst{d-1}}   \qquad
R \ \defi \  1_{\overline{\ga}}\times V \cdot T \qquad
\overline{R} \ \defi \ \eps(\overline{\rho},\rho) \cdot R ,
\]
one can easily checks that
$R$ and $\overline{R}$ solve the conjugate equations
for $\rho$ and $\overline{\rho}$. Now,
let $\ga$ be a simple object of $\Bim_\mathrm{fh}$ localized in $o$.
Since $\ga$ is doubly faithful,
by Proposition \ref{Ac:2}.b   $\ga(1)$ has central support
$1\otimes 1_{n_\ga}$ in $\A(o)\otimes\mathbb{M}_{n_\ga}$. Since
the local algebras are properly infinite,
there exists an isometry
$V:\mathcal{H}_o\longrightarrow\mathcal{H}_o\otimes \mathbb{C}^{n_\ga}$
with values $V_i$ in $\A(o)$ for $i=1,\ldots, n_\ga$, such that
$VV^+=\ga(1)$. So, $\ga_1( \ ) \defi V^+ \ga( \ ) V$ is
automorphism of $\A$ localized in $o$ and transportable. Hence, as in DHR
analysis, it turns out  that $\ga^{\sst{-1}}_1$  is a  conjugate of $\ga$.
\end{proof}
\section{Conclusions}
\label{G}
We have shown that in a tensor $\mathrm{C}^*$-category associated
with a set of representations of a net of local observables
which are local excitations of a reference representation,
the charge structure, in the sense of DHR analysis, manifests itself
even when the Borchers property of
the reference representation is not assumed. What it is essential
is that the local algebras are properly infinite
in the reference representation.\\
\indent  The main problem we have solved, has been to identify
the subcategory carrying the charge structure of the theory,
that is  the subcategory whose objects have conjugates.
Apart from the finiteness of the statistics, the sectors of this category
are characterized by a new property called homogeneity.
The key result allowing us to formulate
this property has been that the theory can be equivalently
studied using both the net and the  presheaf approach.\\
\indent As mentioned at the beginning, the superselection sectors of a net
of local observables on globally hyperbolic spacetimes have
been studied under the assumption that the reference representation
fulfills the Borchers property \cite{GLRV}.
We observed that this assumption has been verified only for certain
models \cite{Ve}. The results here suggest
that it is reasonable to include the proper infiniteness
of the algebras of local observables in the reference representation
as an axiom of the theory. In this case
the charge structure of the theory is carried not by
the subcategory whose objects have finite statistics but by  the one
generated by the homogeneous sectors with finite statistics.
\appendix
\numberwithin{equation}{section}
\section{Some notions and results on tensor $\mathrm{C}^*$-categories}
\label{X}
We introduce the definition of a tensor $\mathrm{C}^*$-category  and
prove some results  concerning left inverses and  simple objects.
The last part of this section is devoted to  exposing some relations
concerning the notation introduced in Section \ref{C}.
The references are \cite{DR1,LR}.\\[8pt]
\noindent \textbf{Left inverses, symmetry  and simple objects }\\[2pt]
Let $\mathcal{C}$ be a category.
We denote by $\rho, \si, \tau,$ etc.. the objects of the category
and the set of the arrows between $\rho,\si$ by $(\rho,\si)$.
The composition of arrows is indicated by ``$\cdot$'' and
the unit arrow of $\rho$ by $1_\rho$. Sometimes, if no confusion is possible,
we will omit the symbol ``$\cdot$'' when we will write the composition
of arrows. \\
\indent $\mathcal{C}$ is said to be a $\mathrm{C}^*$-category if
the set of the arrows between two objects $(\rho,\si)$ is a complex
Banach space and the composition between arrows is bilinear;
there is an adjoint, that is an involutive contravariant functor $*$
acting as the  identity on the objects; the norm satisfies the
$\mathrm{C}^*$-property,
namely  $\norm{R^{*} R} \ = \ \norm{R}^2$ for each
$R\in(\rho,\si)$. Notice, that if $\mathcal{C}$ is a $\mathrm{C}^*$-category
then the set of the form $(\rho,\rho)$ is a $\mathrm{C}^*$-algebra for
each $\rho$. \\
\indent Assume that $\mathcal{C}$ is  a $\mathrm{C}^*$-category. An arrow
$V\in(\rho,\si)$ is said to be  an isometry if $V^* V=1_\rho$;
a unitary, if it is an isometry and $V V^*=1_\si$.
The property of  admitting  a unitary arrow, defines an equivalence
relation on the set of the objects of the category. We denote
by the symbol $[\rho]$ the unitary equivalence class of the object
$\rho$. An object $\si$ is said to be irreducible if
$(\si,\si)=\mathbb{C}1_\si$. $\mathcal{C}$ is said to be closed
under  subobjects if for each orthogonal projection
$E\in(\rho,\rho)$, $E\ne 0$ there exists an isometry
$V\in(\be,\rho)$ such that $V V^{*}  =  E$.
$\mathcal{C}$ is said to be closed under  direct sums,
if given $\rho_i \ i=1,2 $ there exists an
object $\alpha$ and two isometries $W_i\in(\rho_i,\al)$  such that
$W_1  W^{*}_1 \ + \ W_2 W^{*}_2 \ = \ 1_\al$. \\
\indent A \textit{strict tensor $\mathrm{C}^*$-category}
(or \textit{tensor $\mathrm{C}^*$-category}) is a
$\mathrm{C}^*$-category $\mathcal{C}$ equipped with a tensor product,
namely  an associative bifunctor
$\otimes : \mathcal{C}\times\mathcal{C}\longrightarrow\mathcal{C}$
with a unit $\io$, commuting with $*$, bilinear on the arrows
and satisfying the exchange property, i.e.
$(T\otimes S)\cdot (T'\otimes S') \ = \ T T'\otimes S S'$
when the composition of the arrows is defined.
To simplify the  notation we omit the symbol $\otimes$ when
applied to objects, namely $\rho\si \defi \rho\otimes \si$. \\[5pt]
\indent From now on, we assume that
$\mathcal{C}$ is a tensor $\mathrm{C}^*$-category
closed under direct sums, subobjects, and
the identity object $\io$ is irreducible. \\[5pt]
\indent A {\em left inverse} $\Psi$
of an object $\rho$ is a set of nonzero linear maps
$\Psi  =  \{ \Psi_{\si,\tau}:
 (\rho\si,\rho\tau)\longrightarrow(\si,\tau) \}$ satisfying
\begin{align*}
 i) \ \ & \ \Psi_{\si',\tau'}(1_\rho\otimes T\cdot X\cdot 1_\rho\otimes S^{*}) =
  T \Psi_{\si,\tau}(X)  S^{*} &  T\in(\tau,\tau'), \ S\in(\si,\si')\\
 ii) \ &  \ \Psi_{\si\pi,\tau\pi}(X\otimes 1_\pi)   =
   \Psi_{\si,\tau}(X)\otimes 1_\pi & X\in(\rho\si,\rho\tau)
\end{align*}
$\Psi$ is said to be
{\em positive} if $\Psi_{\si,\si}$ is
positive $\forall\si\in\mathcal{C}$; \
{\em faithful} if $\Psi_{\si,\si}$ is positive and
faithful $\forall\si\in\mathcal{C}$;
{\em normalized} if $\Psi_{\io,\io}(1_\rho) \ = \ 1_{\io}$. \\[5pt]
\indent From now on  by   left inverse we mean a positive
normalized left inverse.
\begin{lemma}
\label{Xa:1}
Let $\Psi$ be a left inverse of $\rho$. The following relations
hold:\\
a) $\Psi_{\si,\ga}(R)^* \ = \ \Psi_{\ga,\si}(R^*)$, \ \
   $R\in(\rho\si,\rho\ga)$; \\
b) $\Psi_{\si,\si}(R^* R) \ \geq
 \Psi_{\ga,\si}(R^*)\cdot\Psi_{\si,\ga}(R)$, \ \
$R\in(\rho\si,\rho\ga)$
\end{lemma}
\begin{proof}
a) By polarization of the identity the assertion holds
for the $\mathrm{C}^*$-algebra $(\rho\be,\rho\be)$ for each object $\be$.
For the general case,
let $R\in(\rho\si,\rho\ga)$. Since the category is closed under direct
sums the exists an object $\be$ and
two isometries $V\in(\si,\be)$, $W\in(\ga,\be)$
such that $V V^*  +  W W^*  =  1_\be$.
Since
$\Psi_{\be,\be}(1_\rho \otimes W\cdot R\cdot 1_\rho \otimes V^*)^*$  =
$\Psi_{\be,\be}(1_\rho \otimes V\cdot R^*\cdot 1_\rho \otimes W^*)$,
we have $V\Psi_{\ga,\si}(R)^* W^*  =
 V\Psi_{\si,\ga}(R^*) W^*$ therefore
$\Psi_{\ga,\si}(R)^* = \Psi_{\si,\ga}(R^*)$.
The statement b) follows from the following inequality:
$0  \leq  \Psi_{\si,\si}\left((R \ - \ 1_\rho
    \otimes\Psi_{\si,\ga}(R))^* \cdot
    (R \ - \ 1_\rho \otimes\Psi_{\si,\ga}(R))\right) =$
$\Psi_{\si,\si}(R^* R) -
 \Psi_{\ga,\si}(R^*) \Psi_{\si,\ga}(R)$
\end{proof}
A first consequence of this lemma is the following
\begin{lemma}
\label{Xa:2}
Let $\Psi$ be a left inverse of $\rho$. Then $\Psi$ is zero
if, and only if, $\Psi_{\io,\io}(1_\rho)=0$.
\end{lemma}
\begin{proof}
$(\Rightarrow)$  is trivial.
$(\Leftarrow)$ Let $R\in(\rho\si,\rho\ga)$. Since
$\Psi_{\si,\si}(R^* R) \leq
 \norm{R}^2 \Psi_{\si,\si}(1_{\rho\si}) =
\norm{R}^2(\Psi_{\io,\io}(1_\rho)\otimes 1_\si)$,
we have $\Psi_{\si,\si}(R^*  R) \ = 0$.
By  Lemma \ref{Xa:1}.b  $\Psi_{\si,\ga}(R)=0$.
\end{proof}
Let $\Psi,\Psi^1,\Psi^2$ be,  respectively,
left inverses of $\rho,\rho_1,\rho_2$.
Let $\al$ be the direct sums of $\rho_1$
and $\rho_2$ and let $\be$ be a subobject of $\rho$. Hence there
are two isometries $W_i\in(\rho_i,\al)$ for $i=1,2$
such that $1_\al = W_1 W^*_1 + W_2 W^*_2$ and there is
an isometry $V\in(\be,\rho)$ such that $V V^* \defi E$.
Then the sets $\Psi^1\circ\Psi^2$, $\Psi^1\oplus^s\Psi^2$ for
$s\in [0,1]$, and $\Psi^{\sst{E}}$ defined by
\begin{align}
\label{Xa:2a}
 &(\Psi^1\circ\Psi^2)_{\si,\ga}( \ )  \equiv
  \Psi^1_{\si,\ga}(\Psi^2_{\rho_2\si,\rho_2\ga}( \ ))  \\
\label{Xa:2b}
 &(\Psi^1\oplus^s\Psi^2)_{\si,\ga}( \ )  \equiv
   s  \Psi^1_{\si,\ga}(W^*_1\otimes 1_\ga\cdot
 ( \ )\cdot W_1\otimes 1_\si)   \notag \\
  & \qquad\qquad \qquad\qquad\qquad\qquad\qquad\qquad +  \ (1-s)
 \Psi^2_{\si,\ga}(W^*_2\otimes 1_\ga \cdot
 ( \ )  \cdot W_2\otimes 1_\si)  \\
\label{Xa:2c}
  & \Psi^{\sst{E}}_{\si,\ga}( \ )  \equiv  (\Psi_{\io,\io}(E))^{\sst{-1}} \
   \Psi_{\si,\ga}(V\otimes 1_\ga
  \cdot( \  )\cdot V^{*}\otimes 1_\si)  \qquad \ if \
  \Psi_{\io,\io}(E)\ne 0
\end{align}
are, respectively, left inverses for $\rho_2\rho_1$,  $\al$   and $\be$.
Let us observe that  $\Psi^{\sst{E}}$ is defined if $\Psi_{\io,\io}(E)\ne0$.
Hence, for an object the existence of  left inverses
does not imply the existence of left inverses for its  subobjects. \\[5pt]
\indent A {\em symmetry} $\eps$ in the tensor $\mathrm{C}^*$-category
$\mathcal{C}$ is a map
\[
Obj(\mathcal{C})\ni \rho,\si \longrightarrow\eps(\rho,\si)\in(\rho\si,\si\rho)
\]
satisfying the relations:
\[
\begin{array}{rlrl}
i) & \eps(\rho,\si)\cdot T\otimes S \ = \ S\otimes T\cdot\eps(\tau,\be)
   \ \ \ &   ii) &   \eps(\rho,\si)^* \ = \ \eps(\si,\rho)  \\
iii)&  \eps(\rho,\tau\si) \ = \ 1_\tau\otimes\eps(\rho,\si)\cdot
                     \eps(\rho,\tau)\otimes 1_{\si}  \ \ \ &
iv) &   \eps(\rho,\si)\cdot\eps(\si,\rho) \ = \ 1_{\si\rho},
\end{array}
\]
where $T\in(\tau,\rho), S\in(\be,\si)$.
By $ii)- iv)$ it follows that
$\eps(\rho,\io)=\eps(\io,\rho) = 1_\rho$ for each $\rho$.\\[5pt]
\indent From now on we assume that $\mathcal{C}$ has a symmetry $\eps$.
\begin{prop}
\label{Xa:3}
Let $\Psi$ be a left inverse of $\rho$. Then, \\
$\norm{\Psi_{\rho,\rho}(R^* R) \ \otimes 1_\rho} \ \geq \
 \norm{R\cdot(\Psi_{\rho,\rho}(\eps(\rho,\rho))\otimes 1_\si)}^2 \qquad
 R\in(\rho\si,\rho\ga).$
\end{prop}
\begin{proof}
By using the properties i), iii)  of $\eps$  we have
$R^*  R\otimes  1_\rho$
$ = 1_\rho\otimes\eps(\rho,\si)\cdot\eps(\rho,\rho)\otimes1_\si\cdot
 (1_\rho\otimes R^* R)\cdot\eps(\rho,\rho)\otimes 1_\si\cdot
 1_\rho\otimes\eps(\si,\rho)$.
Using  this relation and Lemma \ref{Xa:1}.a  we have
$\Psi_{\si,\si} (R^* R) \otimes 1_\rho \geq $
$(\eps(\rho,\si)\cdot\Psi_{\rho\si,\rho\si}
 (\eps(\rho,\rho)\otimes 1_\si)\cdot R^* )\cdot
 (R\cdot\Psi_{\rho\si,\rho\si}(\eps(\rho,\rho)\otimes 1_\si))
 \cdot\eps(\si,\rho))$, that  implies  the inequality
written in the statement.
\end{proof}
Let us now recall that given  two left inverses
$\Phi$, $\Psi$ of
$\rho$ and $\si$ respectively, then the following relation holds
(\cite[Section 3.2.6, Lemma 6]{Ro}):
\begin{equation}
\label{Xa:3a}
(\Psi\circ \Phi)_{\rho\si,\rho\si}(\eps(\rho\si,\rho\si)) =
    \Phi_{\rho,\rho}(\eps(\rho,\rho))\times \Psi_{\si,\si}(\eps(\si,\si))
\end{equation}
\begin{prop}
\label{Xa:3b}
Let $\Phi$, $\Psi$ be
two left inverses of $\rho$ and $\si$ respectively. Assume
that $\Psi_{\rho,\rho}(\eps(\rho,\rho))= \la_\rho\cdot 1_\rho $ and
$\Psi_{\si,\si}(\eps(\si,\si))= \la_\si\cdot 1_\si$ where
$\la_\rho,\la_\si\in\mathbb{R}$. If $\la_\rho\ne 0$ and $\la_\si\ne 0$,
then  $\Psi\circ\Phi$ is faithful.
\end{prop}
\begin{proof}
Using the relation (\ref{Xa:3a}) we have
$(\Psi\circ\Phi)_{\rho\si,\rho\si}(\eps(\rho\si,\rho\si))=
\la_\rho\la_\si \cdot 1_{\rho\si}\ne 0$. The proof follows
from Proposition \ref{Xa:3}.
\end{proof}
An object $\ga$ is said to be
{\em simple} if  $\eps(\ga,\ga)=\chi_\ga 1_{\ga^2}$. Since
$\eps(\ga,\ga)$ is self-adjoint and unitary, $\chi_\ga\in\{1,-1\}$.
\begin{lemma}
\label{Xa:4}
Let $\ga$ be simple. Then:\\
a) $\ga^{\sst{n}}$ is simple for each integer  $n$. Moreover,
if $\ga$ has left inverses  then: \\
b) each left inverse of $\ga$ is faithful; \\
c) $\ga^{\sst{n}}$ is irreducible for each integer $n$.
\end{lemma}
\begin{proof}
a) follows from  the property iii) of the  symmetry.
b) follows from  Proposition \ref{Xa:3}.
c) By a) it suffices to  prove the statement for $\ga$.
Let us  assume that  $\ga$ is reducible.
Then there exists an orthogonal projection
$E\in(\ga,\ga)$ such that $E\ne 1_\ga$. Moreover
$0\ne E\otimes (1_\ga - E)\in(\ga^{\sst{2}},\ga^{\sst{2}})$.
In fact, since each left inverse $\Psi$ of $\ga$ is faithful,
$\Psi_{\ga,\ga}((1_\ga - E)\otimes E)= c \  E$ where
$c  1_\io =\Psi_{\io,\io}((1_\ga - E))\ne 0$.
By virtue of this fact we have
$E\otimes (1_\ga - E) = \chi_\ga \
\eps({\ga,\ga})\cdot (E\otimes (1_\ga - E)) =
 \chi_\ga ((1_\ga - E)\otimes E)\cdot\eps({\ga,\ga}) =
 (1_\ga - E)\otimes E$
which gives  rise to a contradiction.
\end{proof}
\begin{prop}
\label{Xa:5}
Let $\Psi$ be a left inverse of $\ga$.
Then the following properties are equivalent:
\[
a) \  \Psi_{\ga,\ga}(\eps(\ga,\ga)) = \pm
1_{\ga} \qquad
b) \ \ga  \ is \ simple  \qquad
c) \  \ga^{\sst{2}} \ is \ irreducible.
\]
\end{prop}
\begin{proof}
a) $\Rightarrow$ b) By Lemma \ref{Xa:3} $\Psi$ is faithful. Since
$(1_{\ga^2} \mp \eps({\ga,\ga}))\in(\ga^{\sst{2}},\ga^{\sst{2}})$ is
positive, we have $1_{\ga^2} =\pm \eps({\ga,\ga})$.
b) $\Rightarrow$  c) follows from the previous lemma.
c) $\Rightarrow$ a) is obvious.
\end{proof}
\noindent\textbf{The notation introduced in Section \ref{C}}\\[2pt]
Given $\rho\in\Dt$,  let us consider, for each pair
$\si,\tau\in\Dt$,
a bounded linear operator
$T\in\mathfrak{B}(\mathcal{H}_o\otimes\mathbb{C}^{n_\rho n_\si},
   \mathcal{H}_o\otimes\mathbb{C}^{n_\rho n_\tau} )$
with values $T_{i,j}$ in $\A$, and
such that $T\rho\si(1) = T = \rho\tau(1)T$.  Such an operator can be
represented as an $n_\tau\times n_\si$ matrix with values $[T]_{i,j}$
in $(\A\otimes\mathbb{M}_{n_\rho})_{\rho(1)}$, that is
\begin{equation}
\label{Xb:3}
 T \ = \ \left(\begin{array}{ccc}
    [T]_{1,1} &  \cdots & [T]_{1,n_\si} \\
    \vdots         &  \vdots & \vdots \\
    {[T]}_{n_\tau,1} &  \cdots & [T]_{n_\tau, n_\si} \\
\end{array}\right) \qquad \mbox{where}\qquad [T]_{i,j}\in (\A\otimes\mathbb{M}_{n_\rho})_{\rho(1)}, \ \
\end{equation}
and $i= 1,\cdots, n_\tau$ and $j= 1,\cdots, n_\si$.
The following relations hold: \\[5pt]
\indent $\bullet$ if $\rho,\si\in\Dt$ then
\begin{equation}
\label{Xb:4}
[\rho\si(A)]_{i,j} \ = \ \rho(\si(A)_{i,j}) \qquad  i,j=1\cdots
n_\si \qquad \forall A\in\A
\end{equation}
\indent $\bullet$ if  $F\in (\rho,\rho), \ E\in (\tau,\si)$ then
\begin{equation}
\label{Xb:5}
[F\times E]_{i,j} \ \ = \ \  F\cdot\rho(E_{i,j}) \qquad
  i= 1,\cdots, n_\tau, \ j= 1,\cdots, n_\si
\end{equation}
\indent $\bullet$ if   $G\in(\rho\tau,\rho\si)$,  $L\in(\rho\si,\rho\be)$
then
\begin{equation}
\label{Xb:6}
[L\cdot G]_{i,j} \ \ = \ \ [L]_{i,k}\cdot [G]_{k,j} \qquad
i= 1,\cdots, n_\tau, \ j= 1,\cdots, n_\be
\end{equation}
\noindent {\small \textbf{Acknowledgments.}
I would like to thank  John E. Roberts 
for helpful discussions and his constant interest in this work.
I am also grateful to him and to Daniele Guido
for their critical comments on the preliminary version
of the manuscript. Moreover, I would like to thank the anonymous referee 
for a careful reading of the manuscript.
Finally, I am grateful to  Ezio Vasselli,
Fabio Ciolli, Gerardo Morsella, Gherardo Piacitelli, and Pasquale Zito
for fruitful discussions,
and to Isabella Baccarelli and Patrick O'Keefe for their precious help.}


\begin{thebibliography}{Enquat 23}
\markboth{Bibliography}{Bibliography}

  \bibitem{Bo}
  H.J.Borchers. {\em A remark on a theorem of B.Misra.}
  Commun. Math. Phys. {\bf 4}, (1967), 315-323.

  \bibitem{DM}
  J.Dixmier, O.Mar\'{e}chal.
  {\em Vecteurs totalisateurs d'une alg\`{e}bre de von Neumann.} (French)
  Commun. Math. Phys. {\bf 22}, (1971), 44-50.

  \bibitem{DHR1}
  S.Doplicher, R.Haag, J.E.Roberts.
  {\em Fields observables and gauge transformations I.}
  Commun. Math Phys. {\bf 13}, (1969), 1-23.

  \bibitem{DHR3}
  S.Doplicher, R.Haag, J.E.Roberts.
  {\em Local observables and particle statistics I.}
  Commun. Math Phys. {\bf 23}, (1971), 199-230.

  \bibitem{DHR4}
  S.Doplicher, R.Haag, J.E.Roberts.
  {\em Local observables and particle statistics II.}
  Commun. Math Phys. {\bf 35}, (1974), 49-85 .


  \bibitem{DR1}
  S.Doplicher, J.E.Roberts. {\em A new duality theory for compact groups.}
  Invent. Math. {\bf 98}, No.1, (1989), 157-218.

  \bibitem{DR2}
  S.Doplicher, J.E.Roberts. {\em Why there is a field algebra with a compact
  gauge group describing the superselection sectors in particle physics.}
  Commun. Math. Phys. {\bf 131}, No.1,  (1990), 51-107.

 \bibitem{Fre}
  K.Fredenhagen:
  {\em Superselection Sectors.}
  Lecture Notes, Hamburg University 1994/1995,
  http://www.desy.de/uni-th/lqp/psfiles/superselect.ps.gz



 \bibitem{GLRV}
  D.Guido, R.Longo, J.E.Roberts, R.Verch.
  {\em Charged sectors, spin and statistics in quantum field theory on
       curved spacetimes.}
  Rev. Math. Phys. \textbf{13}, No. 2, (2001), 125-198.

   \bibitem{Ha}
    R. Haag: 
   {\em Local Quantum Physics.}
    2nd ed. Springer Texts and Monographs in Physics,  1996.

  \bibitem{LR}
  R.Longo, J.E.Roberts.
  {\em A theory of dimension.}
  K-Theory {\bf 11}, No.2, (1997),  103-159.

   \bibitem{Ro2}
   J.E. Roberts.
   {\em  Local cohomology and superselection structure.}
   Commun. Math. Phys. {\bf 51}, (1976), 107-119.

   \bibitem{Ro3}
   J.E. Roberts.
   {\em  New light on the mathematical structure of algebraic field theory.}
   In: {\em Operator algebras and applications, Part2.} (Kingston, Ont. 1980)
   Proc. Sympos. Pure Math. \textbf{38},  523-550, Amer. Math. Soc.,
   Providence, R.I.,  (1982)

   \bibitem{Ro}
   J.E.Roberts.
   {\em Lectures on algebraic quantum field theory.} In:
   {\em The algebraic theory of superselection sectors.}
   (Palermo 1989), D.Kastler ed.,  1-112,
   World Sci. Publishing, River Edge, NJ, 1990.



   \bibitem{Ru}
   G. Ruzzi.
   {\em  Essential properties of the vacuum representation in the
   theory of superselection sectors.} PhD thesis,
    Universit\`a di Genova, March 2002.

  \bibitem{SV}
  K.Szlach\'{a}nyi, P.Vecserny\'{e}s.
  {\em Quantum symmetry and braid group statistics in G-spin models.}
  Commun. Math. Phys. {\bf 156}, No.1, (1993), 127-168.

 \bibitem{Ve}
 R.Verch.
 {\em Continuity of symplectically adjoint maps and the algebraic structure
 of  Hadamard vacuum representations for quantum fields in
       curved spacetime.}
  Rev. Math. Phys. {\bf 9}, No.5,  (1997), 635-674.

  \bibitem{Ve1}
  R.Verch.
  {\em On generalizations of spectrum condition.} In:
  {\em Mathematical Physics in Mathematics and Physics.} (Siena 2000),
   R.Longo ed.,  Fields Ins. Commun. \textbf{30}, 409 - 428,
   Amer. Math. Soc. Providence, RI, 2001

\end{thebibliography}
\end{document}